\documentclass[12pt,preprint]{aastex}

\begin{document}

\title{Abundance Analyses of Field RV Tauri Stars, VI: An Extended Sample  }

\author{
Sunetra Giridhar}
\affil{Indian Institute of Astrophysics;
Bangalore,  560034 India\\
giridhar@iiap.res.in}

\author{David L.\ Lambert }
\affil{The W.J. McDonald Observatory; University of
Texas; Austin, TX 78712-1083
\\ dll@astro.as.utexas.edu}

\author{Bacham E. Reddy\footnote{Visiting Observer, Cerro Tololo
Inter-American Observatory, which is operated by the Association of
Universities for Research in Astronomy, Inc. under contract
with the US National Science foundation.}}
\affil{Indian Institute of Astrophysics;
Bangalore,  560034 India\\
ereddy@iiap.res.in}

\author{Guillermo Gonzalez}
\affil{Department of Physics and Astronomy; Iowa State University; Ames, IA
50011-3160
\\gonzog@iastate.edu    }

\author{David Yong}
\affil{Department of Astronomy; University of
Texas; Austin, TX 78712-1083
\\ tofu@astro.as.utexas.edu}

\begin{abstract}

An abundance analysis is presented and discussed for a sample of 14 RV Tauri
stars. The present abundance data and those from our previous papers and
by other workers are combined in an attempt to further understanding
of the dust-gas separation process which afflicts many RV Tauri
variables. We propose that a star's intrinsic (i.e., initial) metallicity
is given by the photospheric zinc abundance.
Variables warmer that about 5000~K and with an initial metallicity [Fe/H] $\geq$ $-$1 are
affected by dust-gas separation. 
Variables of all metallicities and cooler than about
$T_{\rm eff} \simeq 5000$ K are unaffected by dust-gas
separation. 
The RV Tauri variables show a spread in their C abundances with
the lower boundary of the points in the C versus Zn plane
falling close to the predicted trend for giants after the
first dredge-up. The upper boundary is inhabited by a few
stars that are carbon-rich.
 The O abundances in the mean follow the
predicted trend from unevolved stars in line with the
expectation that photospheric O abundance is unaffected by the
first dredge-up.
 An evolutionary
scenario involving  mass loss by a first ascent or early-AGB red
giant, the primary star of a binary, is sketched.

{\it Subject headings: stars:abundances -- stars:AGB and post-AGB --
stars: variables (RV\,Tauri)}

\end{abstract}

\section{Introduction}

In this series of papers, we have been exploring the chemical compositions of
RV Tauri variables in the Galactic field. This exploration concludes
here with abundance analyses reported for fourteen variables. Of these,
twelve are analysed for the first time, one (LR Sco) was
previously analysed by us before its status as a RV Tauri
variable was appreciated by Lloyd Evans (1999), 
and one (DY Ori) is analysed more thoroughly
than in our earlier attempt.

Beginning with the analysis of the southern RV Tauri
variable IW\,Car (Giridhar, Rao, \& Lambert 1994, Paper I), we have shown that
the atmospheric composition of a  RV\,Tauri star may be  abnormal  in the
sense that the photospheric abundance anomalies are roughly correlated
with the predicted condensation temperature for low pressure gas
of  solar composition. In particular, elements (e.g., Al,
Ca, Ti, and Sc) with the
highest condensation temperatures
 ($\sim
1600$ K) may be seriously underabundant relative to their abundance
expected from the abundances of elements (e.g., S and Zn) of low condensation
temperature
(Gonzalez, Lambert \& Giridhar 1997a,
Paper II; Gonzalez, Lambert \& Giridhar 1997b, Paper III; Giridhar, Lambert
\& Gonzalez 1998, Paper IV; Giridhar, Lambert \& Gonzalez 2000, Paper V).
Our findings were confirmed by independent analyses of
several RV Tauri stars by Van Winckel et al. (1998), Maas, Van Winckel, \&
Waelkens (2002), and Maas (2003).
These abundance anomalies imply that the RV Tauri's photosphere
is deficient in  those elements which condense most readily into
dust grains.   We  refer to the principal
 operation necessary to achieve the  deficiencies as  dust-gas separation.

 If such a separation is to affect the photospheric composition,
 three conditions must be met.
First (condition A), a site for dust formation  must exist near the
star. Two proposed locations compete to meet this
condition  :  the wind off the  star or a
circumstellar or circumbinary disk. Second (condition B), a mechanism must be
identified to separate dust from gas. It has been supposed that
radiation pressure on the dust grains drives them through the
gas and away from the star(s). Third (condition C), the dust-free
gas accreted by the
star must become the dominant constituent of the star's photosphere.
 Here the principal issue is that the photosphere is a part 
 of a convective envelope. In order for the photosphere to assume a
 composition dominated by dust-free gas, the mass of the envelope
 must be small relative to the mass of accreted gas. 

Our earlier analyses have shown that the
severity of the atmospheric abundance anomalies differs from
one RV Tauri to another. These differences presumably are
clues to the circumstances under which the above
conditions are or are not met. Taking them in inverse order,
the following may be noted. Relevant to condition C, the coolest RV Tauri variables
independent of their metallicity are  free of the
abundance anomalies. 
 One interpretation is that these
stars have deep  convective envelopes which dilute 
accreted gas and prevent the appearance of abundance anomalies even when
gas but not dust is accreted (Paper V).
Stars of intrinsically low metallicity such as the variables in
globular clusters (Gonzalez and Lambert 1997; Russell 1998) 
 and some high-velocity variables in the field
appear immune to the effects of a dust-gas separation. This fact is
relevant to
condition B: it is likely a consequence of the
inability of radiation pressure on dust grains to force a separation of
dust from gas when the mass fraction of dust is very low, as it is
in a metal-poor cool environment. Accretion under these circumstances
will not change the surface composition of the star even if it has
a shallow envelope. 
 
With respect to condition A, the extremely metal-deficient A-type
post-AGB stars are evidence that  dust-gas separation does not have to occur
in a stellar wind (Van Winckel 2003).
 The prototypical example is HR 4049
($T_{\rm eff} = 7600$ K) with an extensive infrared excess from
dust but its stellar wind, if it exists, is surely too hot to
be the site of dust formation. Yet, dust-gas separation has
provided a photospheric abundance of [Fe/H] $\simeq -4.8$ for
a star with an initial abundance [Fe/H] $\approx$ $-$0.4.  Van Winckel, Waelkens, \&
Waters (1995) showed that HR 4049 and  other similar metal-deficient stars were
spectroscopic binaries. The dust-gas separation is presumed to occur
in a circumbinary dusty disk.
Superposition of the pulsational velocity variation on an orbital
variation complicates a demonstration that  all RV Tauri stars
affected by dust-gas separation are binaries.
Certainly, several  RV Tauri stars are known to be spectroscopic binaries.
An assessment of the
direct and indirect evidence (Van Winckel et al. 1998) led
Van Winckel (2003) to write
that `binarity may very well be a common phenomenon
among RV Tauri stars'. The question of dust-gas separation
occurring in a wind off a cool star remains open.

An aim of this  paper  was to enlarge the sample of
RV Tauri variables in order to test in more depth
previous deductions about the dust-gas separation processes. We present
abundances for the fourteen  variables  listed in Table 1. 

This new sample particularly increases the representation of high-velocity
RV Tauri stars.

\section{Observations and Abundance Analyses}

The program stars and dates of observation are listed in Table 1
along with the measured radial velocity, the pulsational period, the
Preston spectroscopic type  A, B, or C (Preston et al. 1963), and the
photometric type a or b.  The photometric type b indicates that
the light curve shows a long-term modulation.  

 A majority of the stars were observed with
 the McDonald Observatory's 2.7m Harlan J. Smith reflector with the
CCD-equipped `2dcoud\'{e}' spectrograph (Tull et al. 1995). A spectral
resolving power
 $R = \lambda/\Delta\lambda \simeq  60000$ was
 used and a broad spectral range was covered in a single exposure.
 A S/N ratio of 80-100 over much of the spectral range was 
 achieved. Figure 1 illustrates the quality of typical spectra.

The stars LR Sco and AZ Sgr were observed at CTIO, Chile with the
4m Blanco telescope equipped with a cross-dispersed echelle Cassegrain
spectrograph and a CCD of 2048 $\times$ 2048 pixels. The spectrograph was set to record the
wavelength interval 4900 \AA\ to 8250 \AA\ in 45 orders. Spectral
coverage was complete between these limits. The resolving power
 $R \simeq$ 35,000 was achieved,
as measured from the Th lines in the Th-Ar comparison spectrum.

Spectra were rejected
if they showed line doubling, markedly asymmetric lines, or
strong emission at H$\beta$. (Emission was almost always present
at H$\alpha$.) It is presumed that the spectra not
showing these characteristics represent the atmosphere at a
time when standard theoretical models may be applicable.
This presumption should be tested by analysis of a series of spectra taken
over the pulsational cycle. This remains to be done but in
previous papers we have analysed several stars using spectra taken at
different phases and obtained consistent results.
A striking example was given in Paper III where three observations
of SS\,Gem gave widely different effective temperatures (4750, 5500, and 6500 K)
but similar results for the composition.

The abundance analyses were performed as described in
earlier papers of this series.
 The 2002 version of the spectrum synthesis code MOOG (Sneden 1973)
 was used with ATLAS model atmospheres ( Kurucz 1993). Molecule
  formation was taken into account in computing the line spectrum.
 Hyperfine splitting was considered for the lines of relevant atoms
 and lines ( e.g. Sc, Mn).
 Atmospheric parameters were 
determined in the usual way from the Fe\,{\sc i} and Fe\,{\sc ii}
lines by demanding excitation and ionization equilibrium, and
that the iron abundance be independent of the equivalent width.
Ionization equilibrium is also satisfied for Si, Ti, and Cr:
the mean abundance difference (in dex) between neutral and ionized lines
is $+0.08\pm0.16$ for Si from 11 stars, $-0.02\pm0.16$ for Ti
from 7 stars, and $0.00\pm0.11$ for Cr from 13 stars. 
The adopted parameters listed in Table 2  were used for the    
 full abundance analysis.  Abundances are referred to the solar abundances
given by Lodders (2003, her Table 1). Results are given in Table 3 for the stars
 obviously carrying the signatures of dust-gas separation and
in Table 4 for the other stars.

\section{The Chemical Compositions}

 The chemical composition of a RV Tauri variable may be a blend of
 several signatures. (i) initial composition of the star,
 (ii) the effects of deep mixing during stellar evolution
on the composition, and (iii) the effects of the dust-gas separation,
and (iv) if the RV Tauri stars are the primary of a spectroscopic
 binary, a change of composition may have resulted from mass transferred
 from the companion.

The  initial composition may be anticipated from  abundances of  elements 
thought to be essentially unaffected in the course of evolution
 as single or binary star and also by
dust-gas separation. Here, S and Zn are deemed to qualify as such elements.
Published abundance analyses of main sequence 
stars show that a star's initial composition is generally
predictable to within a small uncertainty from a determination of
the abundance of one element -- see
Goswami \& Prantzos (2000 -- their Figure 7) for a graphical summary.
  We adopt  S and Zn as the
reference elements here, from which we  predict initial abundances of other
 elements to high accuracy.

Although the prehistory  
of RV Tauri variables is unknown in detail,
it may be assumed that they have experienced the first dredge-up which brings
CN-cycled material into the atmosphere. This reduces the C abundance
and increases the N abundance (Iben 1967).
 If RV Tauri variables have  evolved from
the AGB on which they may have experienced thermal pulses and the
third dredge-up, they are expected to be enriched in C  and
possibly also in the $s$-process elements 
(Busso, Gallino and Wasserburg 1999). 
 This last statement will require modification
if a star has accreted substantial amounts of gas from
a companion directly or through transfer from a circumbinary
disk. If the RV Tauri stars are the primary of a binary,
 a change of composition may have resulted from mass transferred
 from the companion.

 When gas  cools and dust grains form, the abundances of the
 elements in the gas phase are reduced below their initial
 abundances. Calculations of the equilibrium distribution of
 an element between dust and gas have been reported by many
 authors and, most recently, by Lodders (2003; her Table~1).
 Principal factors influencing  the gas phase abundances are
 the temperature,  pressure, and initial composition.

 For elements providing major species of grains (e.g., Al and
 corundum  -- Al$_2$O$_3$), one may define a condensation
 temperature (here, T$^\prime_C$) at which the vapor pressure of the grain equals
 the partial pressure of the species in the gas. Below this
 condensation temperature, a species is highly underabundant
 in the gas.
 Other elements, particularly trace elements, are absorbed by
 major species of grains with a degree of absorption which
 may differ from element to element for a given grain type.
 This has led to the concept of a `50\% condensation
 temperature'  -- the temperature at which 50\% of the
 trace element is in the gas and 50\% in grains.
  We adopt the 50\% condensation
 temperatures given by Lodders (2003) in her Table 8 for
 gas of solar composition (Her Table 1)  at a total pressure
 of 10$^{-4}$ bar.  We denote the adopted temperatures by the
 symbol $T_C$. For elements providing the major species of
 grains,  the difference between $T^\prime_C$ and the cooler $T_C$
 is less than 40 K for all but two elements - Si and Ca -
 which we discuss  below (Section 4.1).

 Our assumption is that the effects of dust-gas separation
 on a stellar composition will be revealed as a correlation
 between the underabundance of an element and that
 element's condensation temperature. Obvious approximations
 may invalidate the assumption:  grain formation
 around the RV Tauri stars may not occur under conditions
 of thermodynamic equilibrium; the pressures may differ from one
 formation site to another; the initial composition  may differ,
 as must certainly be the case for the high velocity stars, from
 the assumed solar mix. Yet, the fact is that, in many cases,
 the underabundances are remarkably smoothly correlated
 with $T_C$ (and $T^\prime_C$).

In the following subsections, we present and discuss the compositions
of the stars in Table 1 but leave comment on the
C, N, and O abundances and heavy (Y to Eu) elements to subsequent sections.
Six stars have a composition greatly affected by dust-gas separation
(or another process): UY CMa, EQ Cas, HP Lyr, DY Ori, LR Sco, and
BZ Sct. For the  remainder of the sample, the signature of dust-gas
separation is less distinct and may be absent.

\subsection{UY Canis Majoris}

Our abundance analysis (Table 3) shows clearly
 that UY CMa is a victim of severe dust-gas separation.
This is not unexpected because Preston et al. (1963) assigned it
their spectroscopic class B, and our earlier analyses of Class B
stars found them affected by dust-gas separation. 
Elements of the highest $T_C$ (e.g., Sc and Ti) are underabundant
by a factor of about 200.
The deficiencies [X/H]  are very well
correlated with condensation
 temperature $T_C$ (Figure 2).\footnote{Usual spectroscopic
notation is adopted: [X/Y] = $\log$(X/Y)$_{\rm star}$ -$\log$(X/Y)$_\odot$. }
Judged by the abundances of Na, S, and Zn,
elements of low condensation temperature,
 the initial metallicity corresponds to
[Fe/H] $\simeq -0.4$. A possibly slightly higher metallicity is
suggested by the O abundance with the implication that even Na, S, and Zn
may be slightly depleted. The
high radial velocity  suggests that 
UY CMa belongs to the thick disk. We have not  adjusted
the [X/H] by the small amounts necessary to
account for the fact that a thick disk star of [Fe/H] $\simeq
-0.4$ has a non-solar mix of elements, i.e., [X/Y] $\neq$ 0 (Reddy et al. 2003).

\subsection{RX Capricorni}
 
Inspection of the abundances listed in Table 4 suggests that 
RX Cap
is either a moderately metal-poor thick disk star or a
rather metal-rich halo star, but possessing a normal
composition unaffected by dust-gas separation. 
The inferred initial  iron abundance is [Fe/H] $\simeq$ -0.7 from
consideration of S, Zn, Fe and Ni abundances. 
Within the possible errors and especially considering that
some  abundances are based on a single line, a majority of
the relative abundances [X/Fe] are  as expected. 
We note, for example,
 the positive [X/Fe] for the $\alpha$-elements
Mg, Si, S, and Ti, and the mild deficiency of Mn.
One detects a  suggestion of
dust-gas separation from the low Sc abundance: [Sc/Fe] = $-$0.4 
when [Sc/Fe] $\simeq$ 0.0 is expected, and
just possibly, from Ca which is slightly underabundant relative to
expectation for an $\alpha$-element.

\subsection{EQ Cassiopeiae}

Three spectra of this  variable were available. One was rejected
because some lines showed distinct doubling and He\,{\sc i}
5875 \AA\ was in emission.
The spectrum from 1999, August 19 was crowded
with lines, but unblended lines found with the help of the Arcturus spectrum
are symmetric. The atmosphere at this time was quite cool ($T_{\rm eff}$ =
4500 K). A useful spectrum from 2001, August 19 caught the star at a warmer
phase ($T_{\rm eff}$ = 5300 K). Although H$\alpha$ was in emission, 
unblended lines
were symmetric and deemed suitable for an abundance analysis.  
A radial velocity of $-$158 $\pm$ 1 km s$^{-1}$ was measured from the
1999 August and 2001 August spectra confirming that the star is of
high velocity.
The two spectra provide similar results for the
elemental abundances. Iron abundances from the two
spectra are given in Table 2. In Table 3, we list the
abundances from the 2001 spectrum. Differences in [X/Fe] between the
2001 and 1999 spectra are in the range $-$0.20 to +0.14 with
differences of less than $\pm$0.1 for nine of the 15 elements
common to the two analyses. 

There are what appear to be
signatures of the dust-gas separation (Figure 3) but the
[X/H] versus $T_C$ relation is not as striking and
simple as that for UY CMa. Preston et al.'s (1963) assignment of
the class B? may reflect 
a hint of the unusual
composition differences between the star and RV B stars like
 UY CMa. The  abundances [S/H] $\simeq$
[Zn/H] $\simeq$ $-$0.4 point to an initial [Fe/H] of $-$0.5 to $-$0.4. 
The gross underabundances of Ca and Sc 
([Ca/H] = $-$2.0 and [Sc/H] = $-$3.1) point to dust-gas separation. Yet,  
among the RV Tauri variables analysed by us in this series of
papers, EQ Cas is unique in showing additional  anomalies. 
In particular, the Na abundance is remarkably low:
[Na/Fe] = -0.9 where all other RV Tauri
variables, even those severely affected by dust-gas separation,
 show a positive [Na/Fe].
Furthermore, our estimate for [Na/Fe] may be an upper limit
because the Na\,{\sc i} lines are very weak. The other anomalies
are the  low values of [Cr/H], [Mn/H], and [Cu/H] and the high  value of
[Ti/H] relative to other
elements of similar $T_C$. 
These outstanding anomalies are shown by the two independent
analyses of the star except that  the Cu\,{\sc i} lines were not
 observed in the 1999 spectrum. Rao \& Reddy (2005) show that
 EQ Cas abundances are tightly  correlated with an element's first 
   ionisation potential (FIP). This suggests that a mechanism 
 other than dust-gas separation has affected the photospheric
 abundances.

\subsection{DF Cygni}

Two spectra were acquired but only one had symmetric unblended lines
suitable for an abundance analysis (Table 4).
Line selection was made by reference to the spectrum of
Arcturus, but
given the crowded spectrum, some key elements (e.g., Al and S) proved
undetectable, and many others are represented by no more than one or
two lines. The iron abundance is close to solar. Setting aside elements
represented by just one line, the sole  apparent anomaly is
Sc ([Sc/H] = [Sc/Fe] = -1.0 not 0.0).
 The  Sc underabundance is suggestive
of the onset of dust-gas separation but, as for other stars in Table 4,
Sc (and Ca, in some stars) is the only indicator of this effect. This isolation
of Sc leads one to wonder about other possible explanations, e.g.,
non-LTE effects. 

\subsection{HP Lyr}

This star  came to our attention through
 Graczyk et al.'s  (2002) report of it as `possibly the hottest
RV Tau type object'. These authors estimated the effective
temperature to be about 7700 K. Our spectrum corresponds to
$T_{\rm eff}$ = 6300 K. The radial velocity of -107 km s$^{-1}$
indicates that HP Lyr is  a high-velocity star.
The intrinsic metallicity of the star as assessed from S, Zn, and Na is
slightly sub-solar ([Fe/H] $\simeq$ $-$0.2). We suppose that HP Lyr may belong 
to the thick disk.
The abundances (Table 3 and Figure 4) show the signature of dust-gas 
separation. The correlation between [X/H] and $T_C$ is tight with
almost no scatter in excess of that attributable to the observational
errors in [X/H].

\subsection{TX Ophiuchi}

Most elements (Table 4) have the abundance expected of a normal star
with [Fe/H] = $-$1.2, the measured iron abundance.
This metallicity and the high radial velocity
suggest that the star belongs to the Galactic thick disk.  
One notes the normal [Na/Fe] and [Zn/Fe] but observes that 
Ca, Sc, and Ti,  elements of high condensation temperature, are
apparently underabundant with
[Ca/Fe], [Sc/Fe], and [Ti/Fe]
by  about 0.5, 0.6,  and 0.3 dex, respectively, below  the
expected initial value for a thick disk star of [Fe/H] $= -1.2$.
Aluminum, also of high condensation temperature, however,
has its expected initial abundance.

A comparison with RX Cap reveals that the abundances scale with the
difference in [Fe/H] of 0.4 dex between the two stars.
 The mean difference in [X/H] in the sense TX Oph minus
RX Cap is 0.5 dex from 20 elements with an element-to-element
scatter consistent with the measurement errors. This consistency extends to
Ca, Ti, and Sc.

\subsection{UZ  Ophiuchi}

Two spectra are suitable for an abundance analysis. The spectrum from the
2.7m telescope taken on 2002, June 30 is our primary source.
 An earlier spectrum from
2001, July 13 taken on the 2.1m with the Sandiford Cassegrain echelle
(McCarthy et al. 1995) provided coverage of the interval 4500 \AA\ to
5200 \AA. Since most of the elements were covered in 2002 
 spectrum and abundances derived at two epochs are in good agreement,
 we chose to use only 2002 estimates.
 The radial velocities of $-$95.1 $\pm$ 0.9 km s$^{-1}$ in
2001 July and $-$90.2 $\pm$ 0.8 km s$^{-1}$ in 2002 June
suggest membership of the Galactic thick disk for UZ Oph.
Abundances are given in Table 4.

The composition resembles  that of RX Cap, a star of
similar metallicity and which we
described as  essentially possessing its initial
composition. Relative to RX Cap, 
the mean abundance difference between UZ Oph and RX Cap
is $\Delta$[X/H] = $-$0.15 from 13 elements from
Na to Zn with no element having a $\Delta$ outside the
errors of measurement. The six heavy elements Y, La, Ba, Ce, Nd, Sm
and Eu do provide a significantly different result: $\Delta$ =
$-$0.63: UZ Oph is relatively underabundant in Y to Sm but not
in Eu relative to RX Cap and the expected initial
composition.

\subsection{DY Orionis}

DY Orionis was analysed in Paper II. Our reanalysis uses a
new spectrum providing greater wavelength coverage which enabled us to cover
more elements relative to the earlier study. 
Our new results are in good agreement with
the earlier results.
The star, as noted in Paper II, is severely affected by
dust-gas separation (Table 3 and Figure 5). Particularly
striking is the sharp onset of depletion for elements
with $T_C$ above 1100 K: Na is barely depleted with
 [Na/Fe] of $-0.3$ but Si, Mn and Cr  have [X/H] $\simeq -1.9$.

\subsection{AI Scorpii}

The abundance analysis (Table 4) implies that 
AI Sco's composition is that expected of a star
with the measured [Fe/H] = -0.7. The apparent
anomalies are a mild overabundance of Na and S
and an underabundance of Ca, Sc, and Ti. 
Relative to RX Cap, Na to Zn are slightly overabundant:
$\Delta$ = +0.13 from 13 elements. The heavy
elements -- Y to Nd --  are underabundant:
$\Delta$ = -0.13 from five elements. Europium has a
normal abundance. These small differences between AI Sco and RX Cap
could be attributed to the inevitable observational errors.

\subsection{LR Scorpii}

This star was analysed previously (Giridhar, Rao, \& Lambert
1992) and discussed as a victim of dust-gas separation
in Paper V. Our new analysis is summarized in Table 3.
Maas (2003) has presented an abundance analysis
based on spectra obtained in 2000 and 2001 when  the spectroscopically
estimated parameters  ($T_{\rm eff}$, $\log g$) were
(6250, 0.5) and (5250, 0.0), respectively. His abundances agree
well with ours, especially for [X/Fe].

 Elements  with $T_C \la $ 1400~K  suggest an initial 
iron abundance [Fe/H] = $-$0.2  which with the  low radial velocity
indicates that LR Sco is  a member of the Galactic  thin
disk. Our abundance analysis (Table 3 and Figure 6)
shows that  elements with a condensation
temperature higher than 1600 K are clearly underabundant
 for which [X/H] $\simeq$ $-$0.2 for all elements is expected.

 Our abundances [X/H] are generally in good agreement with the 1992
results obtained from a 1989 spectrum but for Al, Sc, Ti, and Y  our results
differ from the 1992 values by  $-$0.4, $-$0.4, $-$0.7, and $-$1.5 dex, respectively.
 We note that the quartet are  the  elements  with $T_C$ of about 1600~K.
  May one ascribe the differences between our present and previous
results to
a  growth of the  depletion of the  highest $T_C$ elements   
abundance? According to Lodders \& Fegley, the initial
condensates of the five elements are Al$_2$O$_3$,  CaAl$_{12}$O$_{19}$,
Sc$_2$O$_3$, CaTiO$_3$, and Y$_2$O$_3$. This mix precludes a
substantial depletion of Ca.
Calcium with $T_C$ = 1505~K gave the same result ([Ca/H] = $-$0.5) from 
 the 1989 and 2002 spectra. Maas (2003) abundances may be consistent
with the suspected lowering of the abundances of Al, Sc, Ti and Y 
between 1989 and 2002, but a uniform analysis of all available
spectra is desirable. Observations of LR Sco should be continued.

\subsection {AR Sagittarii}

The abundance analysis is summarized in Table 4.
This high-velocity star shows no convincing evidence for
a dust-gas separation. The initial metallicity is
identified as [Fe/H] $\simeq -1.2$. Relative
to TX Oph, a star of the same measured [Fe/H], the mean
$\Delta$ is 0.0 from 14 elements from Na to Zn, and
$-$0.12 from  seven  elements from Ba to Eu. In short, AR Sgr  and
TX Oph are of essentially identical normal composition.

\subsection{AZ Sagittarii}

The star was in a cool phase at the time of observation and
the spectrum is crowded with lines. The abundance
analysis (Table 4) is based on few lines.
The available spectrum
did not go shortward of about 4900\AA\ and, therefore, the
blue and useful Zn\,{\sc i} lines were not recorded. 
Our spectrum
gives the radial velocity as $-$110 km s$^{-1}$. 
This velocity identifies AZ Sgr as a high-velocity
star. 

Judged with respect to a normal star of the measured [Fe/H]
 = $-$1.6, AZ Sgr's anomalies are a high S abundance and a low
Cr abundance: [S/H] is too high by about 0.8 dex and [Cr/H]
is too low by about 0.8 dex. The S abundance is based on weak
lines and is sensitive to the adopted effective temperature.
The Cr abundance appears to be securely based as it is derived
using five neutral Cr lines of moderate strengths. Relative to
TX Oph and AR Sgr, stars with a slightly higher measured
[Fe/H], the S overabundance is less marked, but the Cr
underabundance remains distinctive. Pending acquisition
and analysis of additional spectra of AZ Sqr, preferably 
at a warmer phase, we adopt the conclusion that this 
star is a  metal-poor high-velocity star unaffected by
dust-gas separation. The Cr underabundance is unexplained.

\subsection{BZ Scuti}

This star's composition (Table 3 and Figure 7)
 shows evidence of dust-gas separation.
Sodium, S, and Zn abundances show that BZ Sct's initial
metallicity was close to solar.  These abundances are
based on unblended lines in the spectrum of this warm RV Tauri
star.
Elements
with a condensation temperature higher than about 1200 K are depleted
by a factor of about eight with an
 element-to-element
scatter  comparable to the errors of measurement. There is a hint that
elements with $T_C$ greater than 1600~K
are  more severely depleted.  The extreme depletion
of La is based on a single line.

\subsection{V Vulpeculae}

This star is clearly C-rich.
 The carbon enrichment derived from C\,{\sc i}
lines is also signaled by the presence of C$_2$ Swan bands.
Apart from carbon and possibly sulfur, the composition
of V Vul  resembles closely that of the other
mildly Fe-poor stars in Table 4. The mean $\Delta$, for
example, relative to RX Cap is +0.40 for 11 elements
from Na to Zn, and +0.27 for 4 elements from Y to Eu,
where the [Fe/H] difference is $+$0.4 also.

\section{The compositions of RV Tauri stars}

\subsection{Abundance Anomalies and Condensation Temperatures}

In recent papers on RV Tauri compositions,
abundance versus
condensation temperature plots have been constructed with
values of $T_C$ taken  from Lodders \& Fegley (1998) which are
also 50\% condensation temperatures. Here,
we adopt $T_C$ from Lodders (2003) (see Section 3).
The  effect of a
substitution of the 2003 for the 1998 estimates of $T_C$ is in
most cases a reduction in the scatter of [X/H] about  the mean
trend with $T_C$ for those stars obviously affected by the dust-gas
separation.  In our present sample, this is certainly the
case for UY CMa, HP Lyr, DY Ori, and LR Sco. 
The elements for which replacement of the 1998 by the 2003
$T_C$ estimates  reduces their status as outliers in the
[X/H] versus $T_C$ plot are notably Si, and Ba. In the case of Si,
$T_C$ is lowered  to 1302~K from 1529~K in 1998. The $T_C$ for Ba is raised
from 1162~K in 1998 to 1447~K in 2003.
For other elements measured by us the change in $T_C$ from the 
1998  and 2003 estimates differs by  less than about
50~K, except for Ca and Cu for which the 2003 values are
lower by 130 -- 140 K, too small of a change to affect the
appearance of a [X/H] versus $T_C$ correlation.

It is useful to look behind the $T_C$ estimates at the highest
temperature condensates
expected in equilibrium for solar composition gas. For this exercise,
Lodders (2003) provides the details.
Our discussion considers the development of condensates
as gas is cooled in equilibrium. As long as equilibrium is
maintained, the discussion is readily reworded to
consider the dissolution of condensates as the gas-dust mixture
is heated.

 Inspection of Figures 2, 4, 6, and
7 show that the four most depleted elements are Al, Sc, Ti, and Y
with very similar $T_C$ ($\simeq 1640$K). The initial condensates are
Al$_2$O$_3$ and CaTiO$_3$ leading at a slightly lower temperature
to the major condensates 
CaAl$_{12}$O$_{19}$ (hibonite) and CaTi$_3$O$_{10}$, Ca$_3$Ti$_2$O$_7$,
and Ca$_4$Ti$_3$O$_{10}$ (three forms of calcium titanate). Calcium
titanate may severely deplete gas of Ti but not Ca; the initial
Ca/Ti ratio is about 20. Similarly, if hibonite scours Al from the
gas, little Ca is removed and the initial ratio Ca/Al $\simeq 1$ is
increased to Ca/Al $\simeq 10$. Sc and Y are depleted as Sc$_2$O$_3$
and Y$_2$O$_3$ dissolve in hibonite.

Calcium and Ba are lost to grains at
 a slightly lower $T_C$.
The minor loss of Ca from the gas to hibonite is
enhanced by the condensation of Ca$_2$Al$_2$SiO$_7$ (gehlenite) leading
to Ca/Al $\sim 1$ in the gas with both Ca and Al depleted for $T_C$ $\simeq
$ 1500~K.  Barium is removed from the gas when BaTiO$_3$ dissolves in 
 calcium titanate.
In the 1998 calculations, Ba was identified as dissolving in
CaTiO$_3$ (perovskite), but in the 2003 calculations it was
considered to dissolve into
forms of calcium titanate. It is interesting that although the mode of
transfer of Ca and Ba to
condensates differs their depletion [X/H] in the gas is generally
quite similar for stars severely affected by dust-gas separation.
 (Rare earths  --- La to Eu in our analyses -- dissolve  in
hibonite and titanate.)

Descending the $T_C$ scale, we encounter Mg and Si. Magnesium is removed
from the gas primarily through Mg$_2$SiO$_4$ (forsterite) with
MgAl$_2$O$_4$ (spinel) also contributing. If this were the sole
way to remove Mg and Si from the gas, the Si/Mg ratio in the
gas would swing from Si/Mg $\simeq 1$ to $\simeq 2$. 
The  $T_C$ for Si listed by Lodders \& Fegley (1998)
is the temperature at which 
 gehlenite (Ca$_2$Al$_2$SiO$_7$) forms but, because Ca and Al
are at least an order of magnitude less abundant than Si,
this condensate removes very little silicon from the gas.
At a slightly lower temperature, MgSiO$_3$ (enstatite) condenses and
then one expects Si/Mg $\sim 1$ and similar depletions of Mg and Si.
The $T_C$ taken from Lodders (2003) refers to condensation of
enstatite.

\subsection{Depletion patterns}

The stars in this and earlier papers of our series exhibiting
dust-gas separation seem to form a family with a single outcast.
The main characteristic of the family is an
ordering of depletions [X/H] by the condensation
temperature $T_C$. In our present sample, the extremes
are marked by LR Sco and the pair HP Lyr and DY Ori.
LR Sco would be deemed to have a normal
composition but for depletions ([X/H] $\sim -1$)
of Al, Sc, Ti, and Y, the elements of the highest $T_C$.
At the other extreme, all elements with $T_C \geq 1200$ K
show a depletion increasing with $T_C$ and attaining a
level  [X/H] $\sim -3$ at the highest $T_C$. This is not a single
parameter family: see, for example, the differences in [X/H] between the
extremes HP Lyr and DY Ori.

BZ Sct appears not to show the family's smooth run of
[X/H] with $T_C$, but the several outliers (Cu, K, Sm, and La) in Figure 7 
are based on a single line. If this is deemed a possible reason for the
scatter, one is left with just Mn as showing a slightly discrepant
[X/H]. R Scuti was considered
quite anomalous by Luck (1981) but, in Paper V, we showed that
the underabundant elements, including the heavy elements
analysed by Luck, were  among those with the highest $T_C$.
With due allowance for errors in the abundance determinations,
differences in initial abundances as a function of metallicity
and population class (i.e., thick versus thin disk), and
small familial differences, all analysed stars but one
may be said to belong to the same family bounded by LR Sco and
HP Lyr-DY Ori\footnote{  
Here, we note a good correspondence between R Sct and LR Sco.
It would be useful to extend the analysis of LR Sco to more
heavy elements: Paper V gave [X/H] for nine elements
between Y and Eu but Table 3 includes just two.}.

 The sole non-member of the family is EQ Cas where the
upper envelope to the run of [X/H] with $T_C$ resembles the
pattern shown by HP Lyr, but several elements fall
well below this envelope: notably, Na, Mn, and Cr. 
  Rao \& Reddy (2005) show that the [X/H] for EQ Cas 
 are well correlated  an elements's FIP (see below).

\subsection{Interstellar Depletions}

The abundances for stars affected by dust-gas separation
 correlate quite
 well with those for the
cool interstellar gas towards $\zeta$ Oph (Savage \& Sembach 1996) 
 (Figure 8).
 Depletions of elements in interstellar gas are
attributed to incorporation of the depleted element into and
onto grains. The grains may have formed primarily in denser regions of
interstellar clouds, but, as the detection and analysis of
grains in meteorites demonstrates, some grains were formed in
outflows of red giants, stellar ejecta: supernovae and novae shells
for example.  Grain formation and depletion of atoms in the gas in 
the interstellar medium  cannot be expected to mimic
very closely the convolution of grain formation, depletion of gas in the dusty
reservoir, separation of gas from dust, accretion of the gas, and
mixing of the gas with the stellar atmosphere.  
Despite these caveats, the close correspondence  between the affected
 RV Tauri stars and the
interstellar gas is suggestive of a similarity
between the physical processes involved.
The differences between the stellar and interstellar depletions
 may be    clues to the physical processes at work. 
 
\subsection{ The Initial Fe Abundances}

  On the assumption that S and Zn are unaffected by dust-gas
depletion, their abundances may be taken to be a star's
initial abundances, and, hence, used to provide estimates
of the initial abundance of Fe and  
other elements. Sulfur and Zn are the sampled elements
with the lowest $T_C$ which are considered to be
unaffected by internal nucleosynthesis and mixing.
The run of S and Zn abundances with Fe/H is known for
unevolved stars (Mishenina et al. 2002; Nissen et al. 2004;
Prochaska et al. 2000; Ryde \& Lambert
2004). We are here interested in the abundances of thin and
thick disk stars for which differences 
in [S/Fe] and [Zn/Fe] at a given [Fe/H] are small, say, about 0.1 dex with
the larger values found for the thick disk stars.
In successive figures we show
[Zn/Fe] and [S/Fe] versus [Fe/H] (Figure~9 \& 10), and [S/Zn] versus [Zn/H] (Figure~11). Stars with a
radial velocity of greater than 100 km s$^{-1}$  are termed
 `high velocity' and shown by filled symbols in the figures.

In Figure 9, we show [Zn/Fe] versus [Fe/H] for all of the
RV Tauri variables considered in this and previous papers
in this series. Results for additional RV Tauri variables
are taken from Maas et al. (2002) for RU Cen and SX Cen,
and Maas (2003) for IRAS 09538-7622, 16230-3410, 17038-4815, and
17233-4330. We have added a point for QY Sge (Rao, Goswami, \&
Lambert 2002) which, on circumstantial evidence, was
described as a dust-obscured RV Tauri variable. We elected
not to add other post-AGB stars discussed by Maas (2003). 
Although, these omissions are spectroscopic
binaries and some show effects of dust-gas separation,
available photometry  does not show RV Tauri-like light variations.       
The run of [Zn/Fe] for unevolved
stars corresponds to [Zn/Fe] $\simeq$ 0 for the range
of [Fe/H] covered in Figure 9. Thick disk stars may
have [Zn/Fe] $\simeq +0.1$.

The limit [Zn/Fe] $\simeq 0.0$ represents the lower
boundary to the distribution of the observed
points. (The star with a distinctly negative value of
[Zn/Fe] is DF Cyg with a crowded spectrum and an uncertain
Zn abundance.)
Stars along the lower boundary are those
with an abundance pattern indicating  little or no 
dust-gas separation. Several  but not all
high-velocity  stars are near the boundary. The upper boundary 
to the
data points follows the dashed line. This
corresponds to  approximately [Zn/Fe] = $-$[Fe/H] i.e.,
the trajectory of a star with an initial abundance  [Fe/H] = 0 
and then depleted in Fe to varying degrees but with Zn undepleted. 
On  the assumptions that Fe but not Zn is
depleted and that initially [Zn/Fe] = 0, the observed Zn
 abundance provides the star's
initial Fe abundance [Fe/H]$_0$. 
By inspection, we see from Figure 9 that
the [Fe/H]$_0$ runs from about $+0.1$ to $-1.4$. This is shown
by horizontal dotted line corresponding to [Zn/Fe] $\sim$ 0. 
It is shown by
the figure that the dust-gas separation is not an either-or
effect. 

The run of [S/Fe] versus [Fe/H] (Figure 10) resembles that in 
Figure 9 but with  differences concerning the  upper and lower 
boundaries. The lower boundary to the distribution of
points runs systematically  about 0.2 dex above the run of
[S/Fe] with [Fe/H] for unevolved stars taken from Nissen et al. (2004).
This offset is possibly attributable to systematic 
errors in one or both analyses. 
The dashed line corresponds to [S/Fe] = [Fe/H]. Several stars fall
above this line.
On the assumption that 
the S abundance is unaffected by dust-gas separation, the
upper boundary in Figure 10 implies that 
the initial iron abundance for some stars is about [Fe/H]$_0$ = $+0.4$,
which seems implausible. A possible alternative explanation is that
the S abundances for the RV Tauri stars are overestimated by
about 0.2 to  0.4 dex. These differences between Figures 9 and 10 necessarily
make an appearance in Figure 11 which shows that the [S/Zn] of the
RV Tauri stars fall generally above the expected trend for unevolved
stars taken from Nissen et al. 

In summary, we suggest that the initial [Fe/H] of the
RV Tauri and related stars is obtainable from their measured
Zn abundance and the assumption that [Zn/Fe] = 0 for normal
stars.

\subsection{Carbon and Oxygen Abundances}

Main sequence stars in evolving to red giants develop a
deep convective envelope which brings CN-cycled products to
the surface and, thus, reduces the surface C abundance and
increases the surface N abundance (Iben 1967). Oxygen is expected to be
very little affected. The reduction in the C abundance
is 0.2 to 0.3 dex (Lambert \& Ries 1981). After completion of
He-core burning, material from the He-burning shell is
dredged to the surface to increase the surface C
abundance and also the O abundance. A star at this stage is known
as an asymptotic giant branch (AGB) star. The He-burning
shell may  experience ignition of a neutron source and
synthesis of heavy elements by the 
the neutron-capture $s$-process. By examining the
C, N, O, and heavy element (e.g., Y and Ba) abundances
of the RV Tauri stars and taking due note of the
initial abundances and the possible depletion
of the surface abundances of these elements, one hopes
to specify the evolutionary state of the RV Tauri variables.

In the following discussion, 
[Zn/H] replaces  [Fe/H] as the representative
of a star's initial composition. The initial C and O
abundances as a function of Zn abundance are taken from
measurements of unevolved stars (Nissen et al. 2004; Akerman
et al. 2004). We adopt the LTE abundances given
by Akerman et al. (2004). Their tabulated NLTE abundances for oxygen are
about 0.1 to 0.2 dex smaller than the LTE values. Non-LTE
abundances were not given for carbon.
 We assume that
 the first dredge-up experienced by a red giant reduced
the initial C abundance by 0.25 dex, but left the initial
O abundance undisturbed.

The observed C abundances (Figure 12) lie above the
predictions for a red giant by about 0.4 dex  with a
real star-to-star scatter. Carbon enrichment is suggested. However 
a similar displacement of observed from initial
abundances occurs for [S/Zn]. Since  both the C and the S
abundances are based on high-excitation atomic lines, the 
displacements may arise from similar systematic errors.
 The observed O abundances (Figure 12) follow quite closely
the trend of the initial abundances with [Zn/H].
The observed C/O abundance ratios versus [Zn/H] are shown in
Figure 12.   The lower boundary to the observed ratios
follows the predicted trend for first dredge-up red giants.
 These may be stars which suffered
a reduction of surface C in becoming a red giant but did not
later experience replenishment of C.
Stars with the highest C/O
ratio at a fixed [Zn/H] fall about 0.5 dex above the
predicted trend. This spread probably represents a
real star-to-star spread in C/O. Very few stars are C-rich,
i.e., $\log$ C/O $\geq 0$. 

In summary, the RV Tauri variables have suffered differing 
degrees of surface enrichment of carbon. Some stars, especially
stars of high-velocity, have the C/O ratio   anticipated for fresh
red giants, i.e., no obvious C enrichment occurred following the
first dredge-up. Other stars have a C/O ratio   indicating up to  a
factor of 10 enrichment over the ratio  of a fresh giant.
  Few stars are carbon rich. There is no clear distinction in
Figure 12 between the stars greatly affected by dust-gas separation
and those unaffected.

\subsection{The $s$-process Abundances}

 Mature AGB  stars are   carbon-rich (C/O $>$ 1) with enrichments
 of $s$-process products. Cool carbon stars on the AGB are $s$-process
 enriched by up to one dex (Abia et al. 2002). Post-AGB carbon-rich
 stars are similarly enriched (Reddy et al. 2002; Van Winckel 2003).

 In the case of the RV Tauri stars  unaffected by dust-gas separation,
 $s$-process abundances are directly obtainable from the measured
 abundances with respect to the iron abundance. No obvious enrichments
 are detected. For stars affected by dust-gas separation, the degree of
 $s$-process enrichment must be judged relative to the abundances of
 other elements of a similarly high $T_C$.  
Inspection of Figures 2 through 7 shows that there is no
detectable overabundance of $s$-process elements.
This is true for all other RV Tauri variables from our
series of papers and those analysed by Maas and
colleagues.  Our earlier suggestion that Ba
may be overabundant in AR Pup and DS Aqr deserves
reexamination. Barium, the only $s$-process element
examined in AR Pup, was represented by a single
strong line (Paper II). A similar situation applies to
DS Aqr from Paper IV.  New spectra of broad wavelength
coverage should be obtained and analysed before DS Aqr and
AR Pup are tagged as `s-process enriched'.

In summary, the RV Tauri variables  are unlikely to be descendants
of luminous AGB stars in which $s$-process
products from the He-burning shell have been
dredged to the stellar surface.

\subsection{Boundary Conditions for Dust-gas Separation}

Dust-gas separation is not ubiquitous among RV Tauri stars.
Inspection of the stars unaffected by the separation shows
that their temperature and metallicity domains are  
both well bounded . These bounds are surely clues to
the site of the dust-gas separation and the process by which a
star of anomalous surface composition is created. 

Our sample of RV Tauri variables show  a low and high temperature
boundary to the stars of anomalous composition.
 Stars of all metallicities
 with $T_{\rm eff} \approx$ 5000 K and cooler 
are unaffected by dust-gas separation.\footnote{Among
the stars with $T_{\rm eff} \leq$ 5000 K,
DY Aql appears to defy the above condition.
Our analysis (Paper III) of DY Aql was
based on very few lines; evidence of depletion rests almost
exclusively on a lone Sc\,{\sc ii} line.}
 This is shown by Figure 13.
Our $T_{\rm eff}$ estimates are spectroscopic determinations
from spectra at phases showing well defined symmetric
absorption lines. The spectroscopic T$_{\rm eff}$ may differ with          
 phases (at which spectroscopic analysis is
pursued) for a given star; for example our three spectra of SS Gem gave
 T$_{\rm eff}$ estimates of 4750K, 5500K and 6500K (Paper III).
Effects of dust-gas separation are seen to the high temperature
end of the RV Tauri sample. This boundary is presumably
set by the blue edge of the instability strip. Depletion is
seen in hotter stars (Van Winckel 2003), notably the extremely metal-poor
star HR 4049 with $T_{\rm eff}$ = 7600 K and  with an observed
iron abundance [Fe/H] $\simeq -4.8$. The oxygen abundance
suggests an initial [Fe/H] near -0.5 for HR 4049.

In paper V, we
suggested that the coolest stars were unaffected because the
convective envelope was of sufficient mass to dilute
accreted gas even were it cleansed of dust. Frankowski (2003)
discusses the mass of the convective envelope ($M_{CE}$) of
post-AGB stars by drawing on published numerical calculations.
It is clear that $M_{CE}$ increases with decreasing effective
temperature: for example, $M_{CE}$ at 4000 K, 5000 K, and
6000 K is 0.016$M_\odot$, 0.002$M_\odot$, and 0.001$M_\odot$,
respectively.  
The $M_{CE}$ at 6000 K is some
1000 times the mass of the observable photosphere.
Abundance deficiencies for the high
$T_C$ elements for stars on the hot side of
the boundary may approach a thousandfold, and, in such cases,
 the surface and
convective envelope must
 be  composed of almost undiluted accreted gas.
Although  the increase of $M_{CE}$ to lower 
temperatures may play a role in suppressing the abundance
anomalies, it hardly seems to offer the full explanation.
 Perhaps, cooler variables have stronger
winds which impede  accretion of gas from a circumbinary disk.

The dust-gas separation appears suppressed among stars of a
low initial metallicity. Our sample of RV Tauri stars
suggests that the limiting initial metallicity is [Fe/H] $\approx -1$.
More metal-rich stars are affected, but less metal-rich stars
do not show the abundance anomalies resulting from dust-gas
separation, even if they are hotter than the above
$T_{\rm eff}$ boundary. For stars with [Fe/H] $\ga -$1, there is no
strong evidence that the dust-gas separation efficiency is
 sensitive to the initial metallicity.

In several examples, an affected RV Tauri star currently has
a metallicity well below the limit [Fe/H] of $-$1 for
 occurrence of a dust-gas separation . This is
especially true for hotter stars like HR 4049. 
Our inference from these observations is that it is
the initial and not the current metallicity 
 that is the key factor in establishing
dust-gas separation. Hence,
the dust-gas reservoir is a region with a      
metallicity in excess of [Fe/H] of -1. In a reservoir of low metallicity
 separation of dust and gas  is inhibited and  
 then if material is accreted from the reservoir by the star, the surface
composition is  unchanged.

There is a third boundary condition that may exist: effective
dust-gas separation and accretion of gas but not dust by a star
occurs efficiently only in a binary system. This is spectacularly the case for
the small sample of hot post-AGB stars like HR 4049 for which radial
velocity monitoring has shown them to be single-lined
spectroscopic binaries (Van Winckel et al. 1995). For the RV Tauri
variables, detection of orbital variations in the face of
pulsational variations is a complexity calling for long
campaigns of radial velocity monitoring. Nonetheless, the
number of RV Tauri variables  known to be spectroscopic binaries
is growing as can be seen from Mass (2003).
Van Winckel et al. (1998) have adduced  indirect evidence in support of
the idea that RV Tauri variables are binaries with a circumbinary disk.

Finally, there is the caution suggested by Rao \& Reddy (2005) from
their consideration of the compositions of EQ Cas and CE Vir that showed 
that dust-gas separation may not always be the dominant factor behind abundance
anomalies. We have mentioned the special abundance anomalies of
EQ Cas in Section 3.3. We remarked upon the lack of the usual correlation
between [X/H] and $T_C$.  Rao \& Reddy find that our [X/H] for EQ Cas
are anti-correlated with first ionisation potential, FIP of the elements.
These authors suggest
that the stellar wind preferentially picks up ions over neutral atoms.
At the top of the stellar atmosphere, low FIP elements will be present
as the ions X$^+$ but atoms of the high FIP elements will remain neutral.
The FIP-effect is a minor contributor to the stars for which anomalies exhibit a
strong $T_C$ correlation (Paper IV).

 \section{Concluding Remarks}

 Conversion of the boundary conditions into an  understanding of how RV Tauri
 stars with and without abundance anomalies arise remains elusive.
 Two contrasting scenarios were sketched in Paper V: the stars with
 anomalous abundances are (1) single stars with dust-gas separation
 occurring in the stellar wind (the S hypothesis), and/or (2)
 binary stars with dust-gas separation occurring in a circumbinary
 disk (the B hypothesis).
 The boundary condition on $T_{\rm eff}$ is met by both hypotheses,
 if the deeper convective envelope  for cooler stars acts to
 dilute the accreted gas. The boundary condition on initial
 metallicity seems more readily met by the
 B hypothesis;  the circumbinary disk is comprised of gas ejected
 by either  the RV Tauri's companion and/or the RV Tauri star before
 the onset of accretion and alteration of the photospheric abundances.
Indeed, mass loss induced by a companion may provide not only the circumbinary
gaseous disk but also shifts the primary star off the giant branch
and into the instability strip.
  This boundary condition is not easily satisfied for all stars
 by the S hypothesis; some RV Tauri stars have metal
 abundances less than the boundary [Fe/H] $\simeq -1$.

 The B hypothesis provides a natural link between the RV Tauri
 stars and the A-type extremely metal-deficient stars like
 HR 4049 which are known to be spectroscopic binaries.
 Binary RV Tauri stars are known. A determined campaign
 of radial velocity measurements for RV Tauri stars, especially
 those displaying anomalous abundances, is needed to establish the
 frequency of binaries.
  It should
 be noted too that the sample of HR 4049-like stars is
 presently very small. A search for additional examples and
 a demonstration that they are (or are not!) spectroscopic
 binaries would also be valuable.

 The B hypothesis requires gas to be accreted from the
 circumbinary disk by the star. This accretion, one presumes,
 competes with the wind from a RV Tauri star.
 Direct detection of the infalling
 gas should be sought. On the S hypothesis, it would seem
 that there must be a circulation of gas up to the
 dust-gas separation sites and back down to the photosphere.
 Spectroscopic evidence for this circulation may be
 obtainable from high-resolution spectroscopy over the
 pulsational period of a star with anomalous abundances.

 Finally, we note that the sparsely populated region of the HR
 diagram that is home to
 RV Tauri variables and other luminous post-AGB stars deserves
 continued exploration. Van Winckel's (2003) review gives
 a thorough description and discussion  of the compositions
 of post-AGB stars of which  few have `normal' abundances.
 Our study of RV Tauri stars began with a chance observation of
 IW Car. Perhaps, continued but systematic spectroscopic
 exploration of post-AGB stars will uncover objects
 that will reveal vital clues to the operation of dust-gas
 separation in the RV Tauri and other stars.
 Galactic post-AGB stars offer the best
opportunity for detailed study of the stellar
photosphere and the circumstellar environment.
Although considerably fainter, post-AGB stars in the
Magellanic Clouds offer some advantages, e.g., their
luminosities are obtainable with fair precision.

We thank the referee for a most thorough and constructive review 
 of the paper.
This research has been supported in part by the Robert A. Welch Foundation of
Houston, Texas.

\clearpage
\begin{planotable}{llcccc}
\tablewidth{6in}
\tablenum{1}
\tablecaption{The Program Stars}
\tablehead{\colhead{Star} & \colhead{Date}
& \colhead{ Rad. Vel.}&   \colhead{Period\tablenotemark{a}}&
\colhead{Spec. group\tablenotemark{b}}& \colhead{Phot.
type.\tablenotemark{a}} \\
  \colhead{} & \colhead{} & \colhead{ km s$^{-1}$} & \colhead{(days)}
& \colhead{} & \colhead{}
  }
\startdata
  UY CMa & 1999 Nov 1 & +128 & 114.6 & B & a \nl
  RX Cap & 1999 Nov 2 & -123 & 67.9 & A & \nodata \nl
  EQ Cas & 1999 Aug 19 & -158  & 58.3 & B? & a \nl
                 & 2001 Aug 19 & -158  & 58.3 & B? & a \nl
  DF Cyg & 1999 Nov 1  & -14 & 49.8& A & \nodata  \nl
  HP Lyr & 2002 Nov 14 & -107  & 69.3$^{c}$ & \nodata & \nodata   \nl
  TX Oph & 1999 Aug 19  &-166  & 135.0& A  & \nodata   \nl
  UZ Oph & 2001 Jul 13  & -95  & 87.4 & A  & a  \nl
                 & 2002 Jun 30  & -90  & 87.4 & A  & a  \nl
  DY Ori & 2002 Nov 14 & -8 & 83.4 & B & a \nl
  AI Sco & 2002 Jun 30  &-35   & 71.0 & A  &  b   \nl
  LR Sco & 2002 Jun 22  &-18   & 104.4 & \nodata  & \nodata   \nl
  AR Sgr & 1999 Oct 31 & -112 & 87.9 & A?  &  a  \nl
  AZ Sgr & 2002 Jun 19 & -110 & 113.6 & \nodata & \nodata   \nl
  BZ Sct & 1999 Jun 6   & +89 & 75.7 &\nodata  & \nodata   \nl
  V Vul  & 1999 Aug 19 & -28  & 81.1 & A &a    \nl
\enddata
\tablenotetext{a}{Data GCVS Kukarkin et al. 1958: Kholopov et al. 1985}
\tablenotetext{b}{data from Preston et al. 1963}
\tablenotetext{c}{Data from Graczyk et al. 2002}
\end{planotable}
\clearpage

\begin{planotable}{lllrlcrlr}
\tablewidth{7in}
\tablenum{2}
\tablecaption{Stellar Parameters Derived from the
Fe-line Analyses}
\tablehead{
    &  & \colhead{Model\tablenotemark{a}} &
  \colhead{$\xi^{\rm b}_{\rm t}$} & \multicolumn{2}{c}{Fe
I\tablenotemark{c}} & &
\multicolumn{2}{c}{Fe II$^{\rm c}$}   \\ \cline{3-3} \cline{5-6} \cline{8-9}
Star &\colhead{~~UT Date} & \colhead{T$_{\rm eff}$, $\log g$,
[Fe/H]} & \colhead{(km~s$^{-1}$)} & \colhead{$\log \epsilon$} &
\colhead{n} & &\colhead{$\log \epsilon$} & \colhead{n}
}
\startdata
UY CMa  &1999 Nov 1 & 5500, 0.0, $-$1.4 & 5.0\phantom{000} & $6.15 \pm 0.17$ &
65 && $6.20 \pm 0.14$ & 11  \nl
RX Cap &1999 Nov 02& 5800, 1.0, $-$0.8 & 4.8\phantom{000} & $6.69 \pm 0.16$ &
80 && $6.67 \pm 0.13$ &  16   \nl
EQ Cas & 1999 Aug 19 & 4500, 0.0 , $-$0.8 & 4.6\phantom{000} & $6.68 \pm 0.13$ &
54 && $6.71 \pm 0.12$ &13   \nl
   & 2001 Aug 19 & 5300, 0.7 , $-$0.8 & 3.6\phantom{000} & $6.73 \pm 0.15$ &
53 && $6.71 \pm 0.14$ & 9   \nl
DF Cyg &1999 Nov 01&  4800, 1.7, $-$0.0 & 3.6\phantom{000} & $7.50 \pm 
0.17$ &22 && $7.36 \pm 0.18$ & 7   \nl
HP Lyr & 2002 Nov 14&  6300, 1.0, $-$1.0 & 3.2\phantom{000} & $6.48 \pm 0.12$ &
99 && $6.52 \pm 0.10$  &15  \nl
TX Oph & 1999 Aug 19& 5000, 0.5, $-$1.2 & 4.9\phantom{000} & $6.24 \pm 0.17$ &
81 && $6.28 \pm 0.16$  &17  \nl
UZ Oph &2001 Jul 13 & 5000, 0.5, $-$0.7 & 3.6\phantom{000} & $6.78 \pm 0.14$ &
68 & &$6.80 \pm 0.10$ &  6   \nl
UZ Oph & 2002 Jun 30 & 4800, 0.0, $-$0.8 & 3.6\phantom{000} & $6.70 \pm 0.15$ &
  98& &$6.63 \pm 0.14$ & 14   \nl
DY Ori & 2002 Nov 14& 6000, 1.5, $-$2.0 & 4.0\phantom{000} & $5.23 \pm 0.14$ &
30 && $5.27 \pm 0.12$  &12  \nl
AI Sco & 2002 Jun 30 & 5300, 0.25, $-$0.7 & 3.3\phantom{000} & $6.76 \pm 0.14$ &
38 & &$6.82 \pm 0.10$ &  8   \nl
LR Sco & 2002 Jun 22 & 6000, 0.50, $-$0.2 & 4.2\phantom{000} & $7.29 \pm 0.16$ &
50 & &$7.24 \pm 0.11$ &  7   \nl
BZ Sct & 1999 Jun 6 & 6250, 1.0, $-$0.8 & 3.2\phantom{000} & $6.65 \pm 0.12$ &
63 && $6.64 \pm 0.13$ &  16  \nl
AR Sgr & 1999 Oct 31 & 5300, 0.0, $-$1.5 & 5.2\phantom{000} & $6.16 \pm 0.13$ &
62 && $6.09 \pm 0.14$ & 17    \nl
AZ Sgr & 2002 Jun 19& 4750, 0.5, $-$1.6 & 4.2\phantom{000} & $4.90 \pm 0.15$ &
45 && $4.93 \pm 0.13$ &  10  \nl
V Vul & 1999 Aug 19& 4500, 0.0, $-$0.4 & 4.6\phantom{000} & $7.10 \pm 0.11$ &
52 && $7.05 \pm 0.11$ &  10  \nl
\enddata
\tablenotetext{a}{$T_{\rm eff}$ in K, log $g$ in cgs, [Fe/H] in dex.}
\tablenotetext{b}{$\xi_{\rm t}$ is the microturbulence determined
from the Fe I lines}
\tablenotetext{c}{$\log \epsilon$ is the mean abundance relative to H
(with $\log \epsilon_{\rm H} = 12.00$).
  The standard deviations of the means, as calculated
from the line-to-line scatter, are given. $n$ is the number of
considered lines. }

\end{planotable}

\clearpage
\tabletypesize{\scriptsize}
\begin{deluxetable}{lccccccccccccc}
\tablecaption{Elemental Abundances}
\tablenum{3.1}
\tablehead{
& &\multicolumn{3}{c}{UY Cma} && \multicolumn{3}{c}{EQ Cas} &&\multicolumn{3}{c}{HP Lyr} \\
\cline{3-5}  \cline{7-9}  \cline{11-13}  \\
\colhead{Species} &\colhead{$\log \epsilon_{\odot}^{o}$}&
\colhead{[X/H]}   &\colhead{ n} &\colhead{[X/Fe]} &&\colhead{[X/H]} &
\colhead{n} &\colhead{[X/Fe]} &&\colhead{[X/H]} &\colhead{n} &\colhead{[X/Fe]} \\
}
\startdata
C I& 8.39 & $-0.03\pm 0.19$& 5& $+1.26$ && $-0.23\pm0.19$&3& $+0.52$&
&$-0.29\pm0.12$&15& $+0.69$\\
N I& 7.83 & &... && && ...               &&&$-0.19\pm0.01$&3& $+0.79$ \\
O I& 8.69 & $-0.18\pm 0.09$& 3& $+1.11$ && $+0.04\pm0.04$& 2& $+0.79$&
& $+0.03\pm0.19$& 4& $+1.01$ \\
Na I& 6.30 & $-0.44$& 1& $+0.85$&&  $-1.65\pm0.13$& 2&
$-0.90$ && $-0.17\pm0.08$&  4& $+0.81$\\
Mg I& 7.55 & $-1.35\pm 0.12$& 4& $-0.06$&&  $-0.52\pm0.19$& 3&
$+0.23$ && $-0.86\pm0.15$& 4&$+0.12$ \\
Al I& 6.46 & $-1.84$& 1& $-0.55$&&  &...&  &&
$-3.21\pm0.09$&  2&$-2.23$ \\
Si I& 7.54 & $-0.97\pm 0.13$& 4& $+0.32$&&  $-0.26\pm0.10$&11&
$+0.49$ && $-0.45\pm0.12$&  6& $+0.53$ \\
Si II& 7.54 & $-0.95$& 1& $+0.34$&&  $-0.29\pm0.23$& 2&
$+0.46$&&  $-0.58\pm0.20$&  2& $+0.40$\\
S I& 7.19 & $-0.32\pm 0.12$& 5& $+0.97$&&  $-0.35\pm0.15$& 3&
$+0.40$&&  $+0.05\pm0.15$&  8& $+1.03$ \\
K I& 5.11 & &... &&  && ...
&&&  & ...         \\
Ca I& 6.34 & $-1.65\pm0.22 $& 4& $-0.36$ && $-2.02\pm0.11$& 4&
$-1.27$ && $-1.95\pm0.14$&  4& $-0.97$ \\
Sc II& 3.07 & $-2.22\pm 0.15 $& 5& $-0.93$&&  $-3.06$& 1&
$-2.31$ && $-2.87$& 1& $-1.89$\\
Ti I& 4.92 & &...      &&  & $-1.32\pm0.17$& 2&
$-0.57$& &&& ...                      \\
Ti II& 4.92 & $-2.38\pm 0.13 $& 5& $-1.09$ && $-1.19\pm0.09$& 7&
$-0.44$ && $-2.97\pm0.19$&  5& $-1.99$ \\
V  I& 4.00 & & ...  && & & ...  &&&  & ...  \\
Cr I& 5.65 & $-1.88\pm 0.11 $& 7& $-0.59$&&  $-2.05\pm0.03$& 4&
$-1.30$ && $-1.24\pm0.16$& 5& $-0.26$ \\
Cr II& 5.65 & $-1.72\pm 0.09 $& 4& $-0.43$ && $-1.90\pm0.10$& 3&
$-1.15$ && $-1.31\pm0.08$& 8& $-0.33$ \\
Mn  I& 5.50 & $-1.12\pm 0.18 $& 5& $+0.17$&&  $-1.78\pm0.09$& 3&
$-1.03$ && $-0.74\pm0.10$&  4&$+0.24$ \\
Fe   & 7.47 & $-1.29$ & &&&$-0.75$ &&&& $-0.98$ & & \\
Co  I& 4.91 && ... & && $-0.59\pm0.06$& 2& $+0.19$ &&& ...&  \\
Ni  I& 6.22 & $-1.57\pm 0.10 $& 6& $-0.28$ && $-0.79\pm0.17$&20&
$-0.04$&&  $-1.14\pm0.12$& 10& $-0.16$ \\
Cu  I& 4.26 & $-0.75\pm 0.20 $& 2& $+0.54$&&  $-1.04$&1&
$-0.29$&&  $-0.47\pm0.02$&  2& $+0.51$ \\
Zn  I& 4.63 & $-0.59\pm 0.06 $& 3& $+0.70$ && $-0.33\pm0.17$& 3&
$+0.46$ && $-0.35\pm0.07$& 4& $+0.63$ \\
Y  II& 2.20 & $-2.32\pm 0.13 $& 4& $-1.03$ && $-1.94\pm0.14$& 5&
$-1.19$ && $-2.78$&  1& $-1.80$ \\
Ba II& 2.18 & $-2.02\pm 0.15 $& 3& $-0.73$ &&& ... &&&
$-1.93\pm0.10$& 2& $-0.95$ \\
La II& 1.18 & &... && && ... & && &...& \\
Ce II& 1.61 & &... && && ... && & &...& \\
Pr II& 0.78 & &... && && ... && & &...& \\
Nd II& 1.46 & &... && && ... && & &...& \\
Sm II& 0.95 & &... && && ... && & &...& \\
Eu II& 0.52 & $-1.31$& 1& $-0.02$& && ... &&& & ...&\\
\enddata
\end{deluxetable}
\clearpage
\tabletypesize{\scriptsize}
\begin{deluxetable}{lccccccccccccc}
\tablecaption{Elemental Abundances}
\tablenum{3.2}
\tablehead{
& &\multicolumn{3}{c}{DY Ori} && \multicolumn{3}{c}{LR Sco} &&\multicolumn{3}{c}{BZ Sct} \\
\cline{3-5}  \cline{7-9}  \cline{11-13}  \\
\colhead{Species} &\colhead{$\log \epsilon_{\odot}^{o}$}&
\colhead{[X/H]}   &\colhead{ n} &\colhead{[X/Fe]} &&\colhead{[X/H]} &
\colhead{n} &\colhead{[X/Fe]} &&\colhead{[X/H]} &\colhead{n} &\colhead{[X/Fe]} \\
}
\startdata
C I& 8.39 & $+0.10\pm0.16$& 16& $+2.33$  & &$-0.05\pm0.13$& 12&
$+0.22$&& $+0.05\pm0.12$& 20&$+0.87$  \\
N I& 7.83 &$+0.54\pm0.01$& 3& $+2.77$&&& \nodata&  &&
$+0.12\pm0.01$&3&$+0.93$ \\
O I& 8.69 &$-0.20\pm0.14$& 2& $+2.43$ &&  $-0.05\pm0.15$& 4& $+0.12$&
& $+0.06\pm0.15$& 4& $+0.88$\\
Na I& 6.30 & $-0.25\pm0.06$& 3 & $+1.98$ & & $-0.04\pm0.17$& 4&
$+0.21$ & & $-0.05\pm0.05$& 4& $+0.77$\\
Mg I& 7.55 & $-2.32\pm0.11$& 2& $+0.09$ & & $-0.29\pm0.08$& 2&
$-0.12$ & & $-0.75\pm0.02$& 3&
$+0.07$ \\
Al I& 6.46 & $-3.29~~~~~~~~~$& 1& $-1.06$ & & $-0.81\pm0.17$&
3&$-0.64$ & && \nodata
&&\\
Si I& 7.54 & $-1.44~~~~~~~~~$& 1& $+0.79$ & & $+0.07\pm0.10$&10&
$+0.24$&  & $-0.59\pm0.01$&
2& $+0.23$ \\
Si II& 7.54 & $-1.69\pm0.03$& 2& $+0.54$ & & $-0.08~~~~~~~~~$&1 & $+0.09$ & &
$-0.73\pm0.15$& 2& $+0.09$ \\
S I& 7.19 & $+0.20\pm0.06$& 8& $+2.43$&  & $+0.15\pm0.05$& 4& $+0.32$ & &
$+0.18\pm0.15$& 5& $+1.00$ \\
K I& 5.11 & & \nodata &&&&\nodata& && $-0.73~~~~~~~~~$& 1& $+0.09$\\
Ca I& 6.34 & $-2.18\pm0.14$& 2 & $+0.05$ && $-0.51\pm0.11$& 6&
$-0.34$ & & $-0.91\pm0.10$& 7& $-0.09$\\
Sc II& 3.07 & $-2.85~~~~~~~~~$& 1 & $-0.62$ && $-1.30\pm0.08$& 4&
$-1.13$ & & $-1.13\pm0.17$& 5& $-0.31$\\ 
Ti I& 4.92 & &\nodata &&& $-0.96~~~~~~~~~$& 1& $-0.79$ & &
$-1.25\pm0.06$& 2& $-0.43$  \\
Ti II& 4.92 & $+2.33\pm0.29$& 6& $-0.10$&& & \nodata & &&
$-1.18\pm0.12$& 6& $-0.36$  \\
V  I& 4.00 && \nodata && &&\nodata && && \nodata  & &\\
Cr I& 5.65 & $-1.93\pm0.06$& 2&$+0.33$& & $-0.28\pm0.05$& 2& $-0.11$&
& $-0.74\pm0.10$& 5& $+0.08$\\
Cr II& 5.65 &$-1.99\pm0.17$& 8& $+0.24$& & $-0.11\pm0.03$& 4& $+0.06$
& & $-0.91\pm0.08$& 9& $-0.09$ \\
Mn  I& 5.50 & $-1.84\pm0.19$& 3& $+0.37$ && $-0.32\pm0.05$& 2&
$-0.15$ & & $-1.09\pm0.13$& 4& $-0.27$\\
Fe   & 7.47 & $-2.23~~~~~~~~~ $& & &&$-0.17~~~~~~~~~ $ &&&&
$-0.82~~~~~~~~~ $ && & \\
Co  I& 4.91 && \nodata &&& &\nodata && & &\nodata&& \\
Ni  I& 6.22 & $-2.28\pm0.26$& 2& $-0.05$& & $-0.46\pm0.17$& 9&
$-0.29$ & & $-0.75\pm0.11$&11& $+0.07$ \\
Cu  I& 4.26 & &\nodata &&& &\nodata && & $-0.75~~~~~~~~~$& 1& $+0.07$\\
Zn  I& 4.63 & $-0.17\pm0.16$& 4& $+2.06$ && $-0.32~~~~~~~~~$& 1&
$-0.15$ & & $+0.04\pm0.05$& 3& $+0.94$\\ 
Y  II& 2.20 & $-2.56\pm0.14$& 3& $+0.33$ && $-1.63\pm0.16$& 2&
$-1.46$ & & $-1.36\pm0.08$& 4& $-0.54$\\ 
Ba II& 2.18 & $-2.05\pm0.12$& 2& $-0.18$ &&& \nodata && &
$-0.91~~~~~~~~~$& 1& $-0.09$ \\
La II& 1.18 & &\nodata & &&&\nodata && & $-1.79~~~~~~~~~$& 1& $-0.97$\\
Ce II& 1.61 & &\nodata &&&& \nodata && & $-0.92\pm0.01$& 2& $-0.10$\\
Pr II& 0.78 & &\nodata &&& & \nodata &&& & \nodata &&\\
Nd II& 1.46 & &\nodata && && \nodata && & &\nodata & \\
Sm II& 0.95 & & \nodata &&& & \nodata && & $-0.61~~~~~~~~~$& 1&
$+0.21$ \\
Eu II& 0.52 & &\nodata &&& $-0.24~~~~~~~~~$& 1& $-0.07$ & &
$-0.64~~~~~~~~~$& 1& $+0.18$\\
\enddata
\end{deluxetable}
\clearpage

\begin{figure}
\plotone{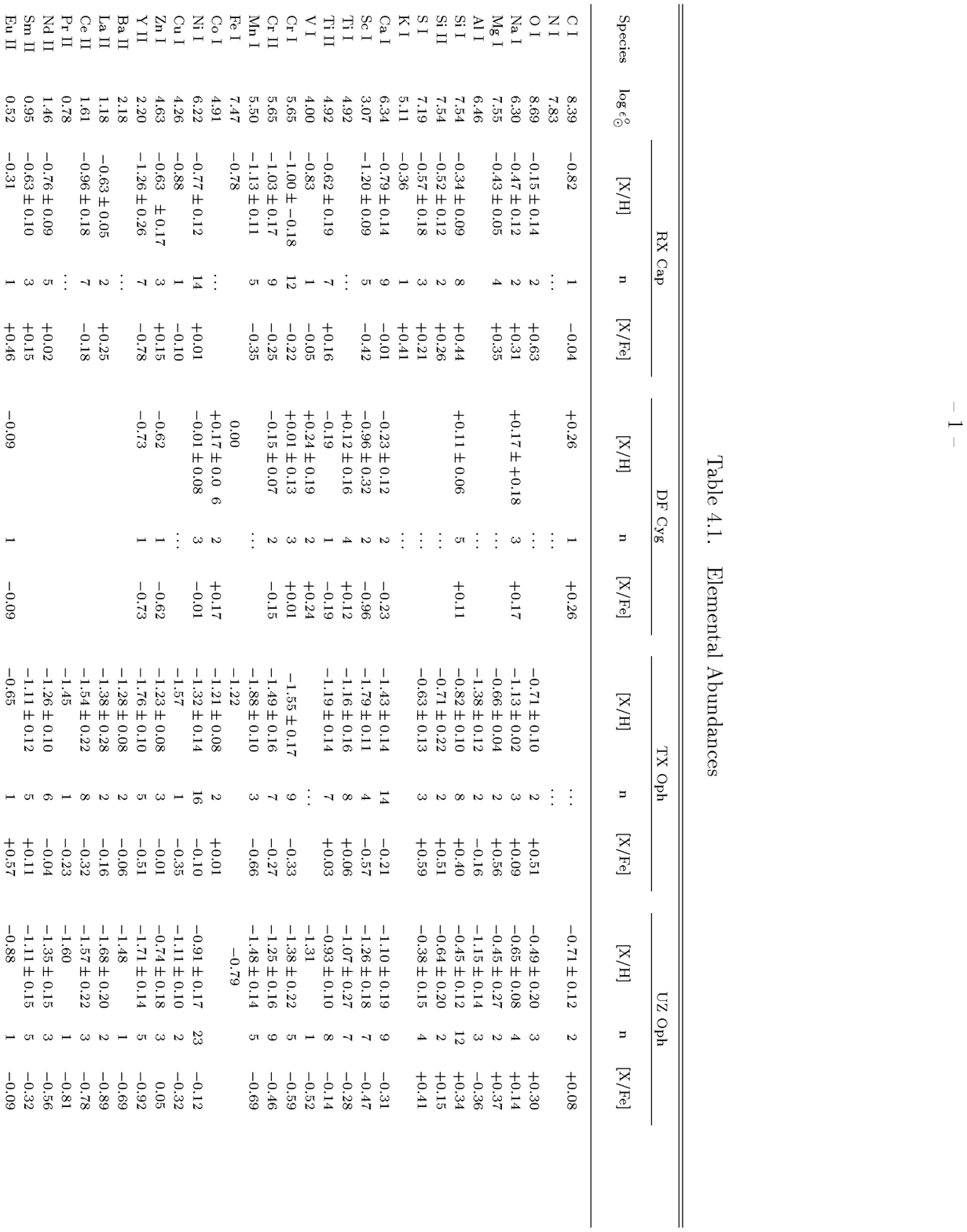}
\end{figure}
\clearpage
\begin{figure}
\plotone{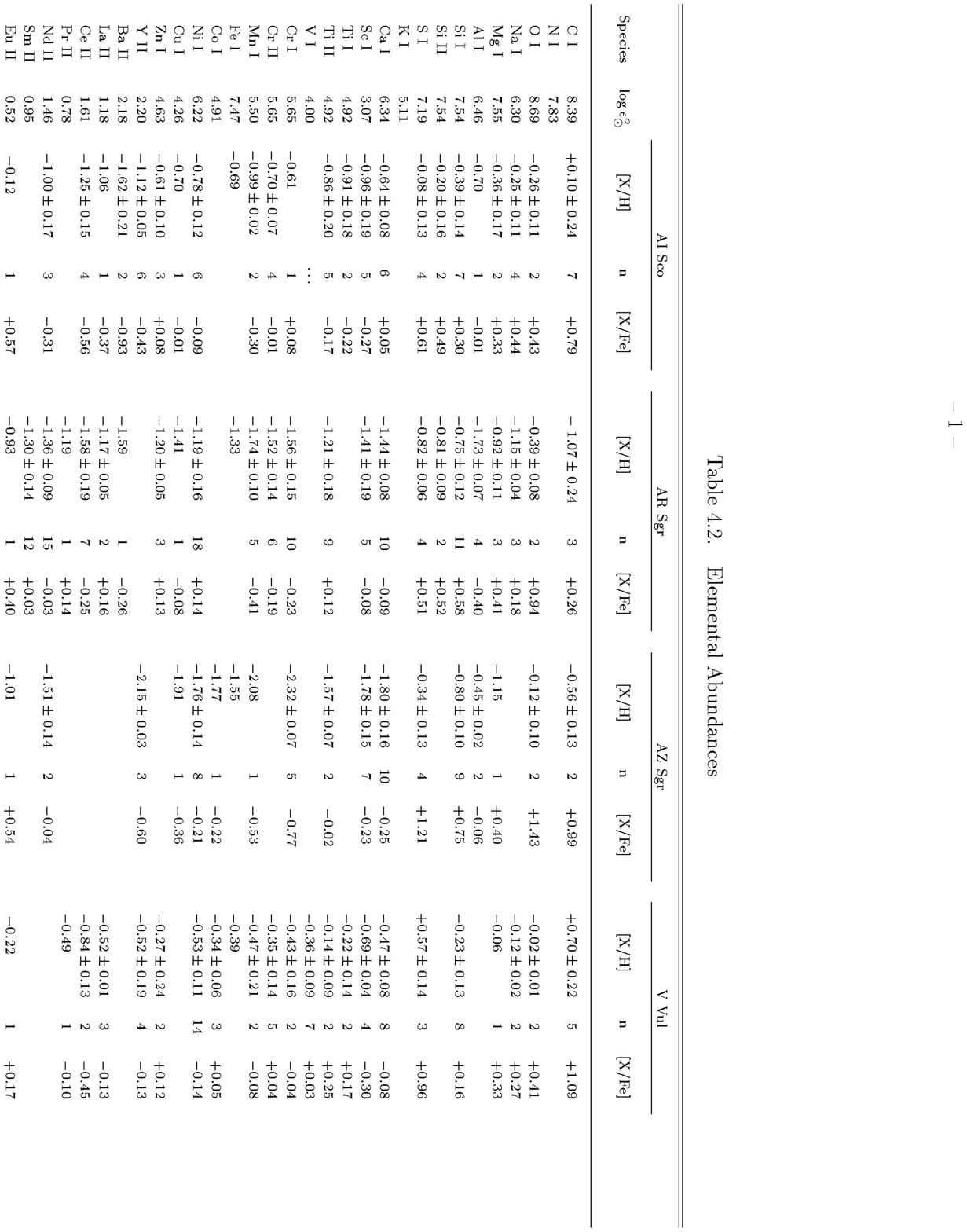}
\end{figure}


\begin{figure}
\plotone{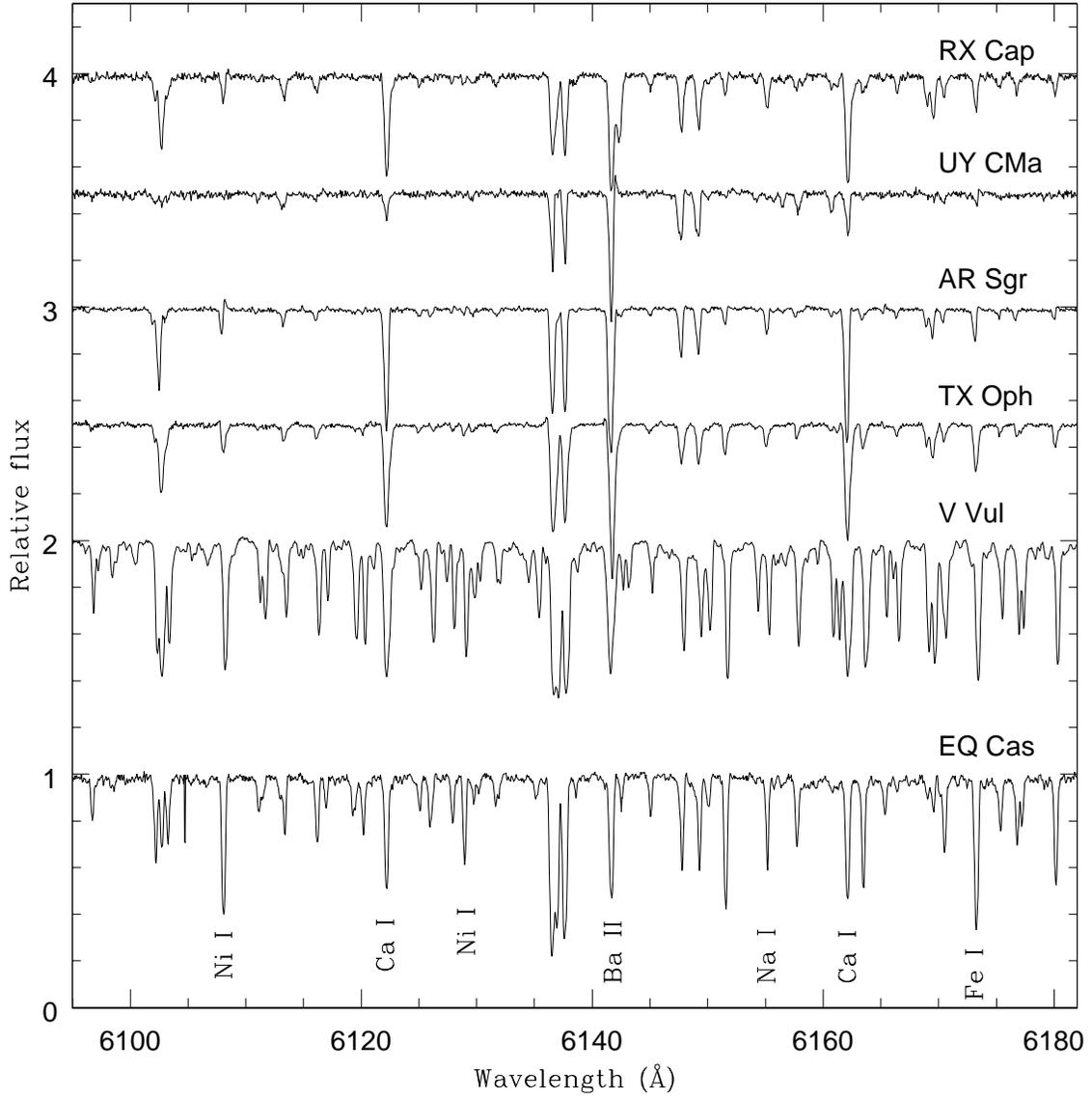}
\caption { Spectra of representative sample  are
presented in descending order of their temperatures (from top to bottom). The temperature of
UY CMa is within 200K of RX Cap and AR Sgr but the lines of Ca I and
Fe I are considerably weak in UY CMa due to dust-gas separation effect.}
\end{figure}

\clearpage

\begin{figure}
\plotone{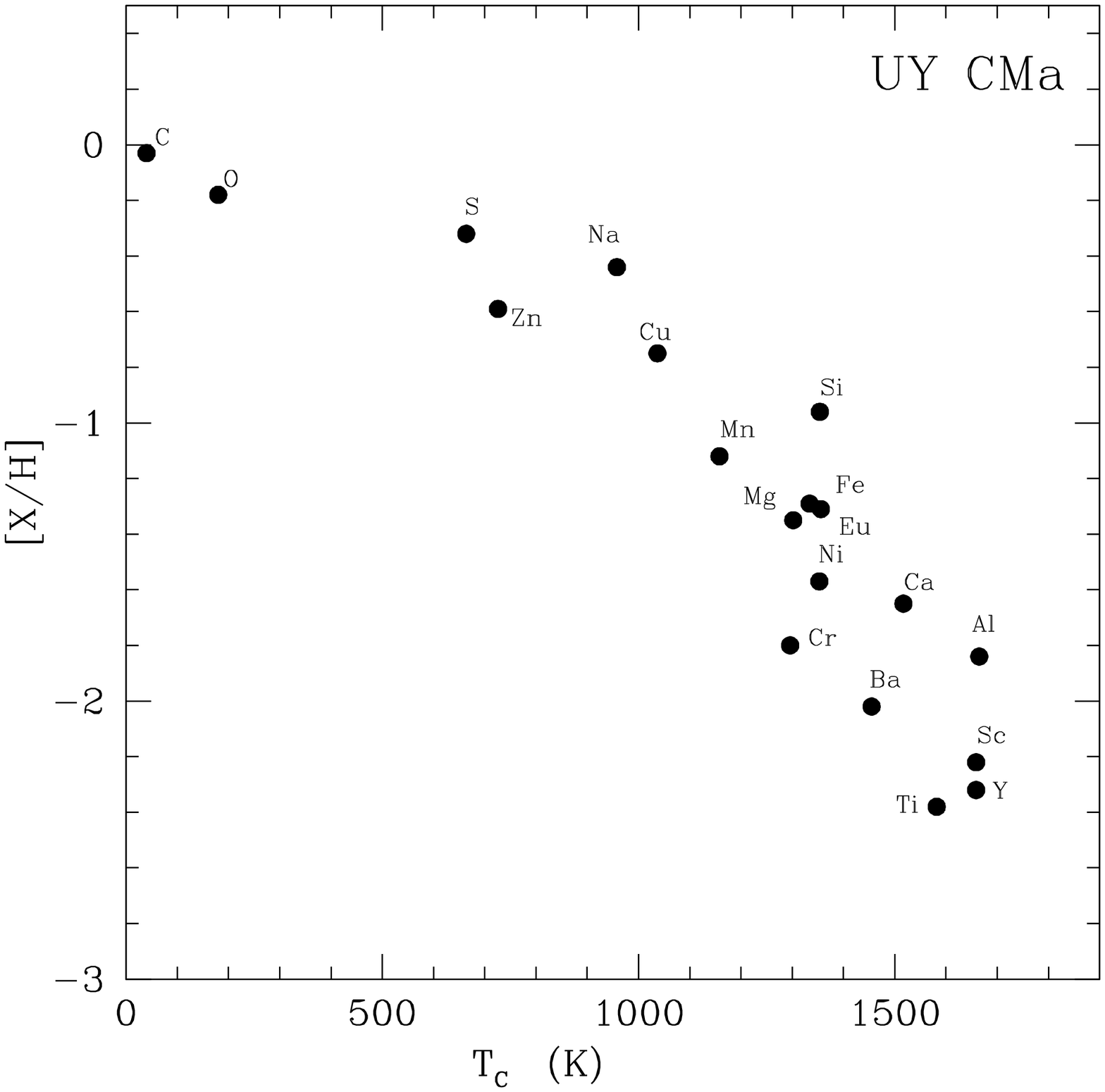}
\caption{Abundance [X/H] versus condensation
temperature $T_C$ for UY\,CMa. Elements are identified by their chemical
symbol.}
\end{figure}

\clearpage

\begin{figure}
\plotone{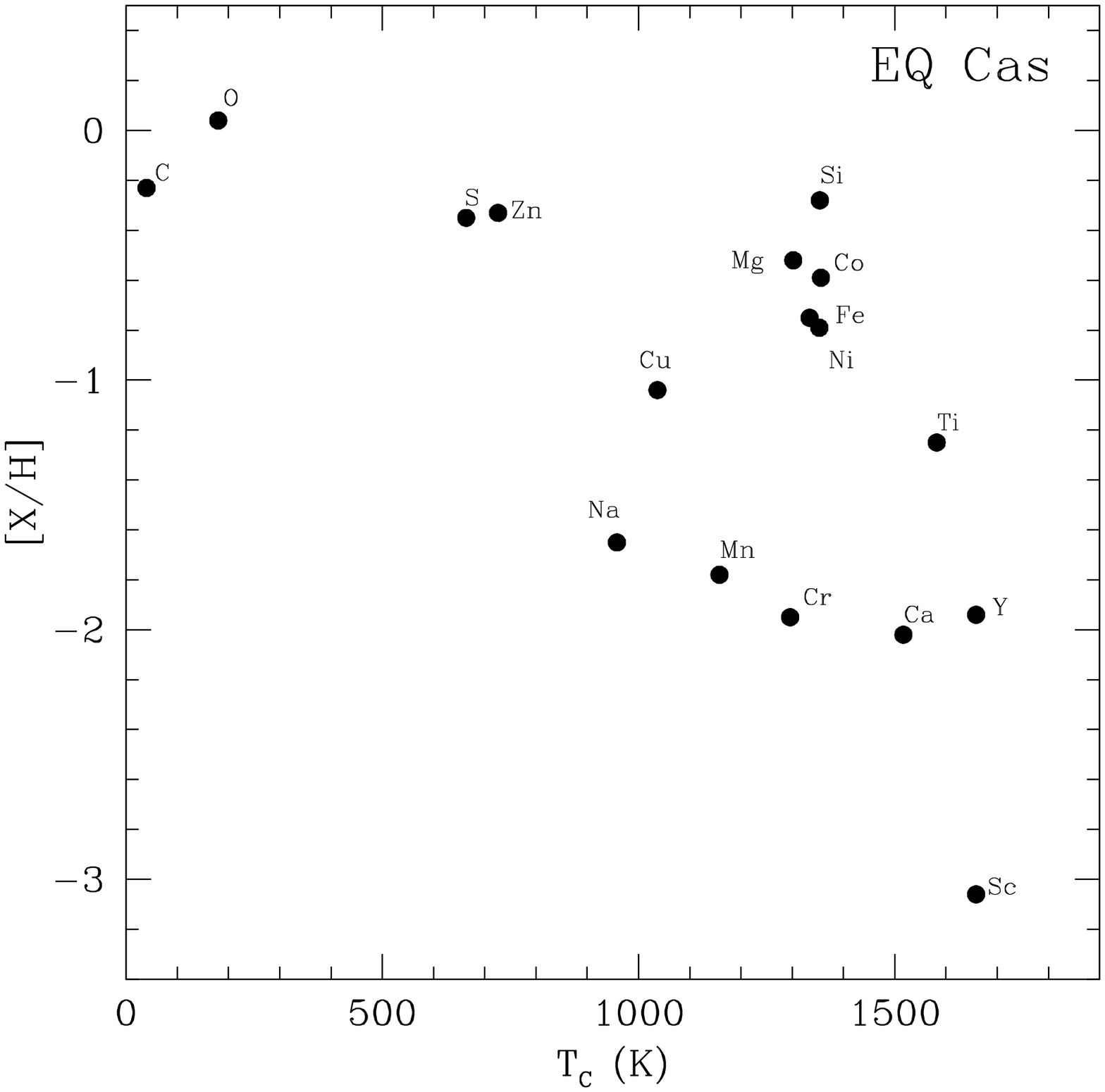}
\caption{Abundance [X/H] versus condensation temperature $T_C$ for
EQ Cas. Elements are identified by their chemical symbol.}
\end{figure}
\clearpage

\begin{figure}
\plotone{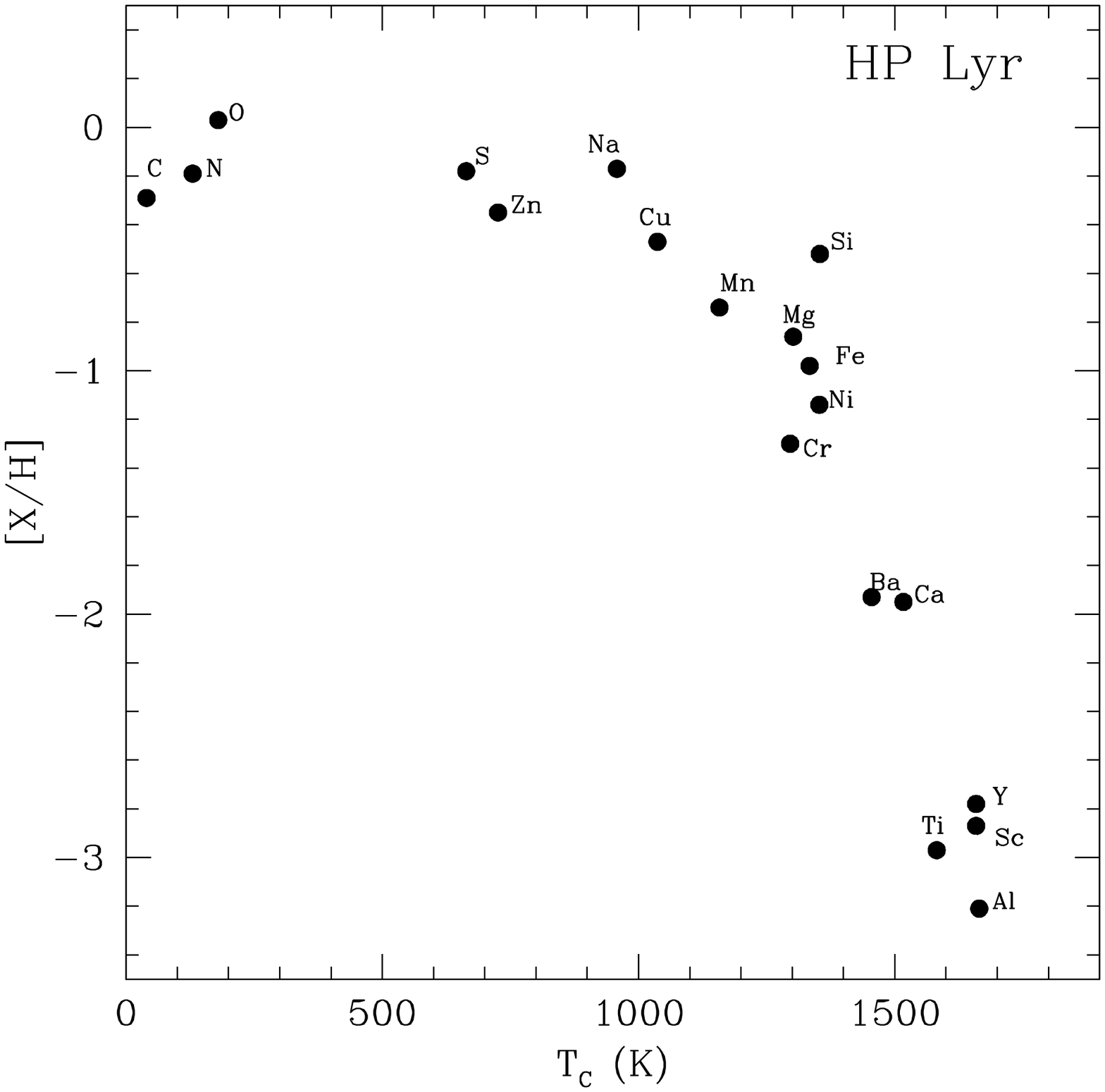}
\caption{Abundance [X/H] versus condensation temperature $T_C$
for HP Lyr. Elements are identified by their chemical symbol.}
\end{figure}
\clearpage

\begin{figure}
\plotone{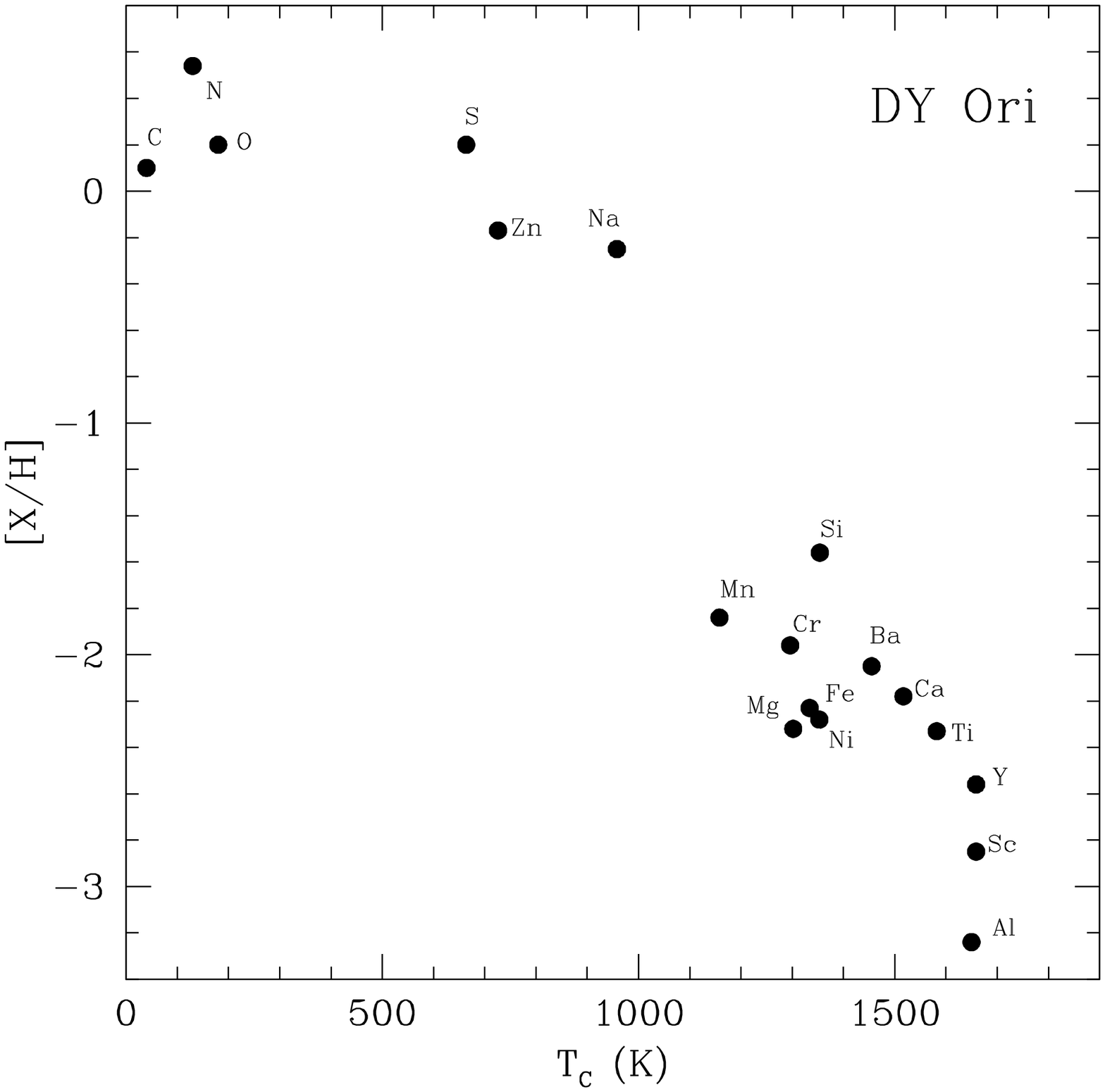}
\caption{Abundance [X/H] versus condensation temperature $T_C$
for DY Ori. Elements are identified by their chemical symbol.}
\end{figure}
\clearpage

\begin{figure}
\plotone{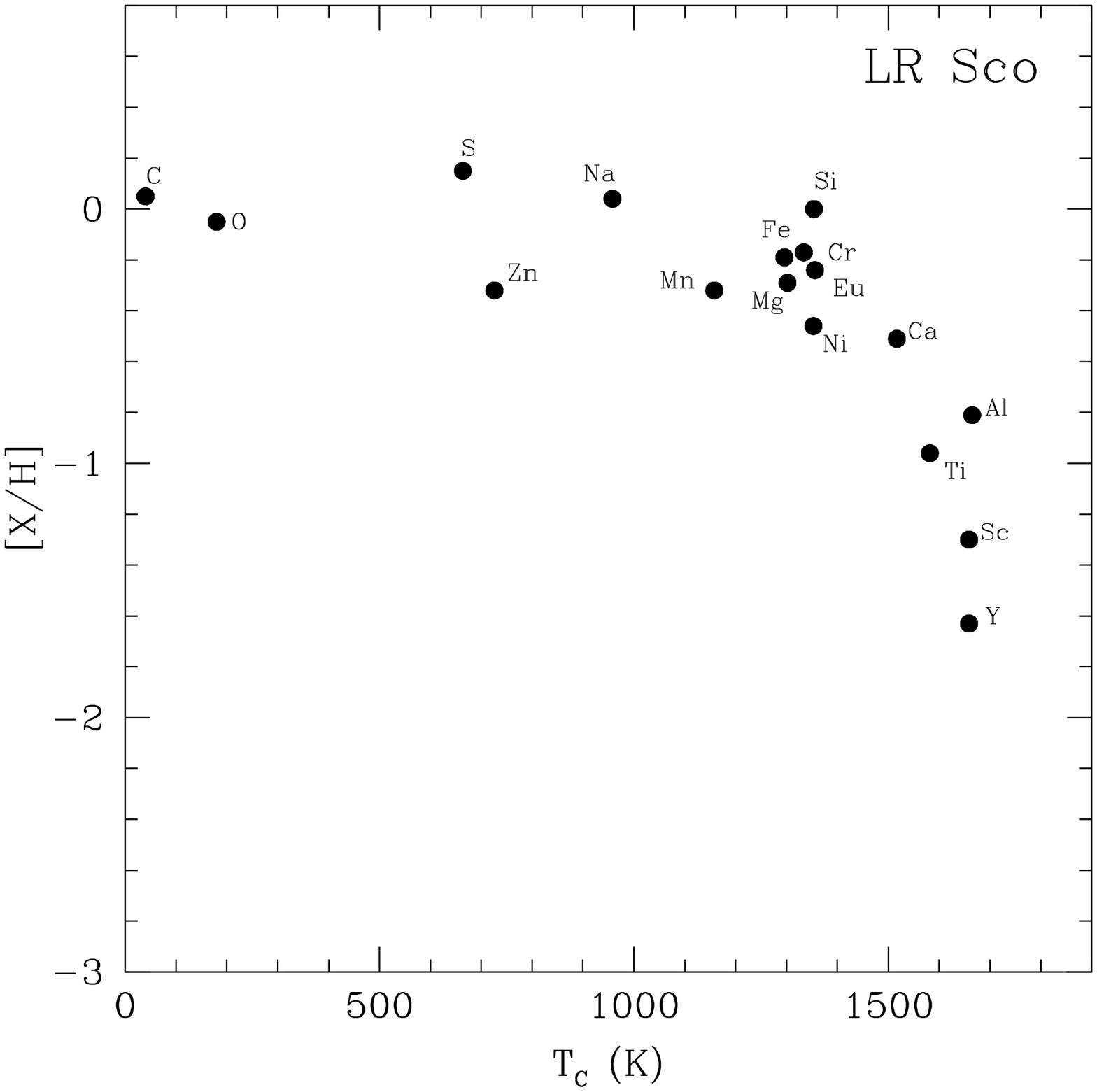}
\caption{Abundance [X/H] versus condensation temperature $T_C$
for LR Sco. Elements are identified by their chemical symbol.}
\end{figure}
\clearpage

\begin{figure}
\plotone{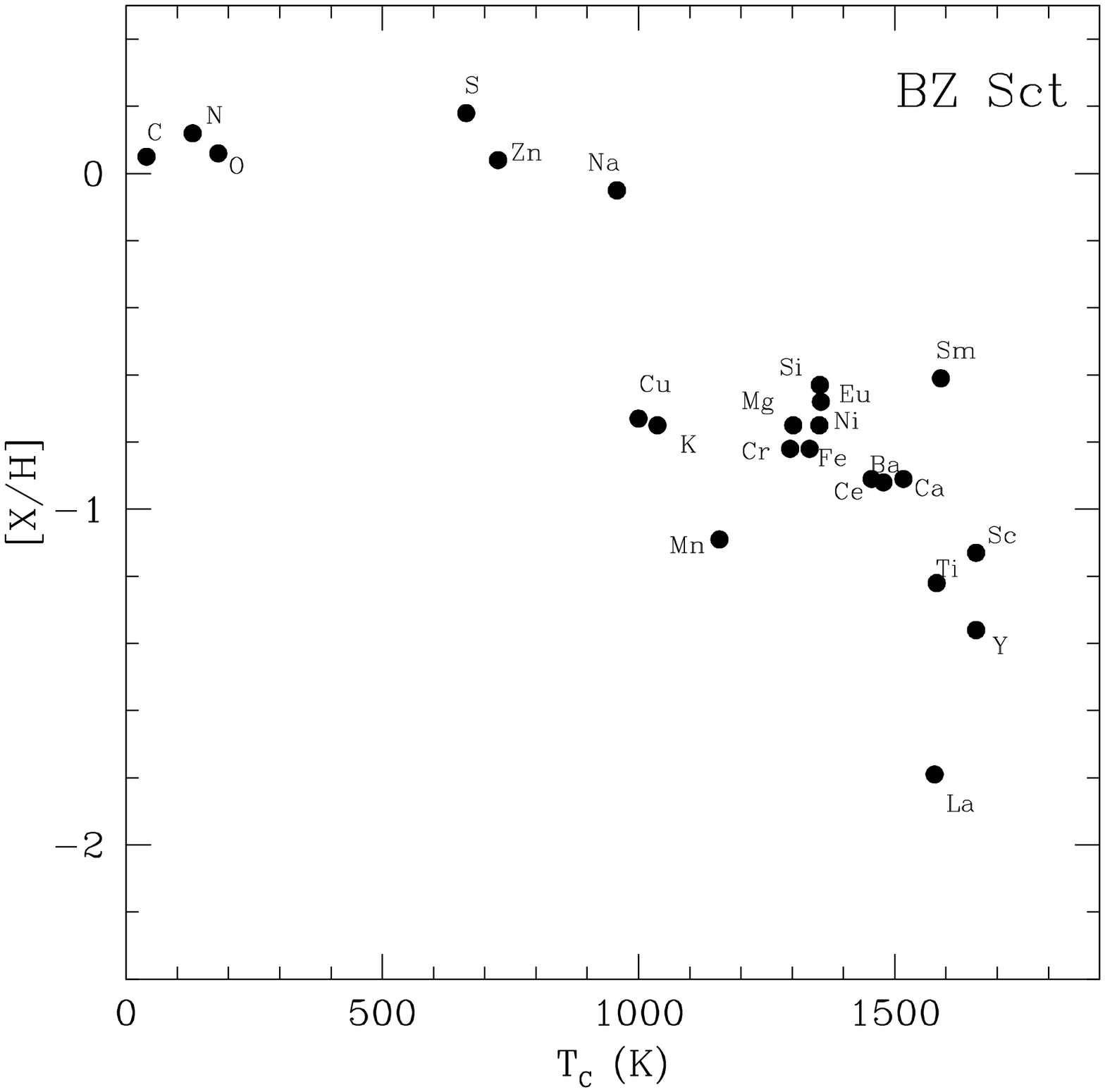}
\caption{Abundance [X/H] versus condensation temperature $T_C$
for BZ Sct. Elements are identified by their chemical symbol.}
\end{figure}
\clearpage

\begin{figure}
\plotone{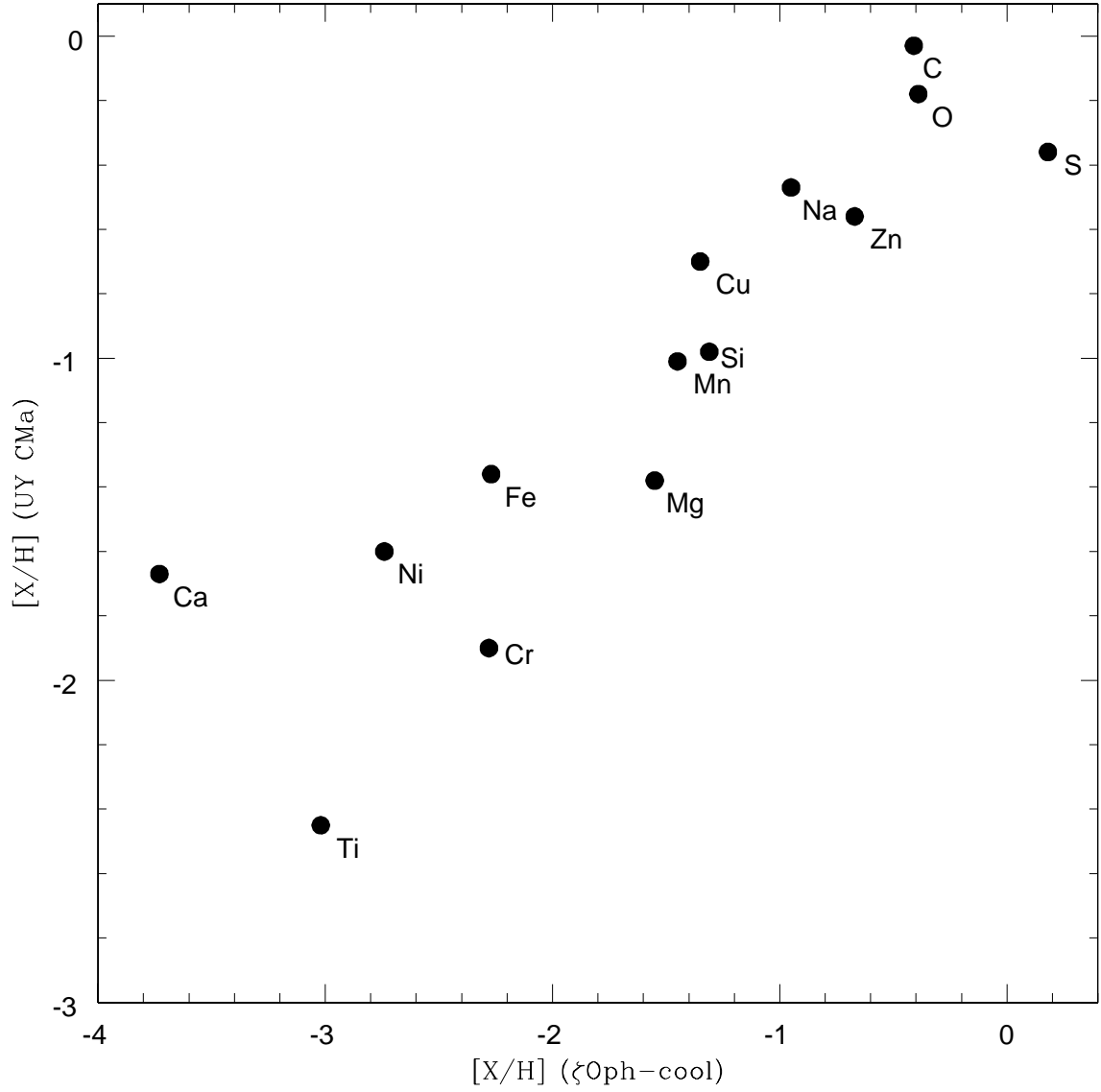}
\caption{Abundance [X/H] for UY CMa versus [X/H] for the cool
interstellar cloud along the line of sight to $\zeta$ Oph.
Elements are identified by their chemical symbol.}
\end{figure}
\clearpage

\begin{figure}
\plotone{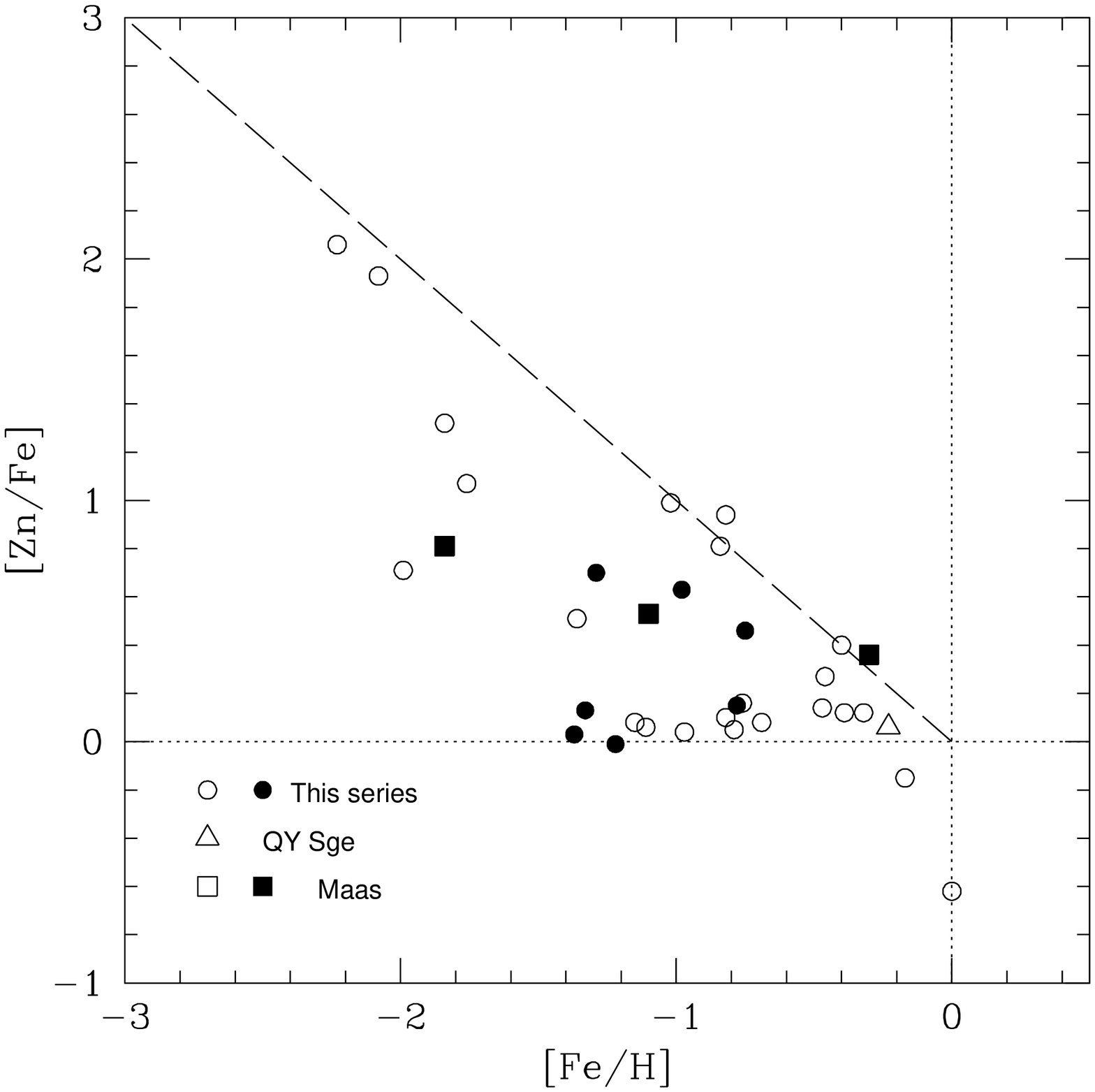}
\caption{Abundance ratio [Zn/Fe] versus abundance [Fe/H] for
RV Tauri and related variables. RV Tauri stars from our series of papers
are represented by  circles. QY Sge (Rao et al. 2002) is shown by
the triangle. RV Tauri stars analysed by Maas et al. (2002) and
Maas (2003) are
shown by squares.
Filled circles and squares denote high velocity stars.
} 
\end{figure}
\clearpage

\begin{figure}
\plotone{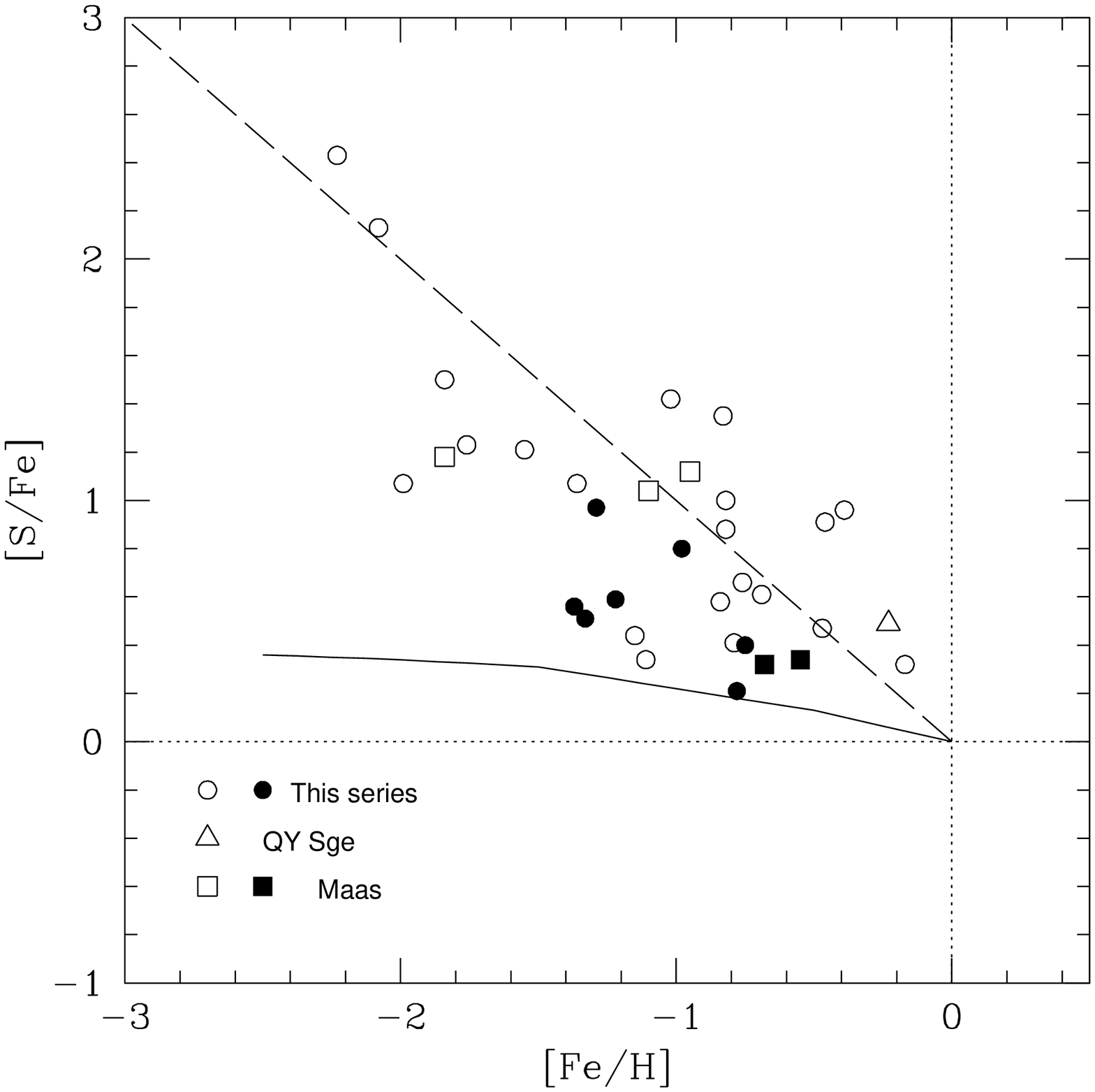}
\caption{Abundance ratio [S/Fe] versus abundance [Fe/H] for
RV Tauri and related variables. RV Tauri stars from our series of papers
are represented by  circles. QY Sge (Rao et al. 2002) is shown by
the triangle. RV Tauri stars analysed by Maas et al. (2002) nd
by Maas (2003) are
shown by squares.
Filled circles and squares denote high velocity stars.
The solid line  shows the run of [S/Fe] with
[Fe/H] for unevolved stars (see text). } 
\end{figure}
\clearpage

\begin{figure}
\plotone{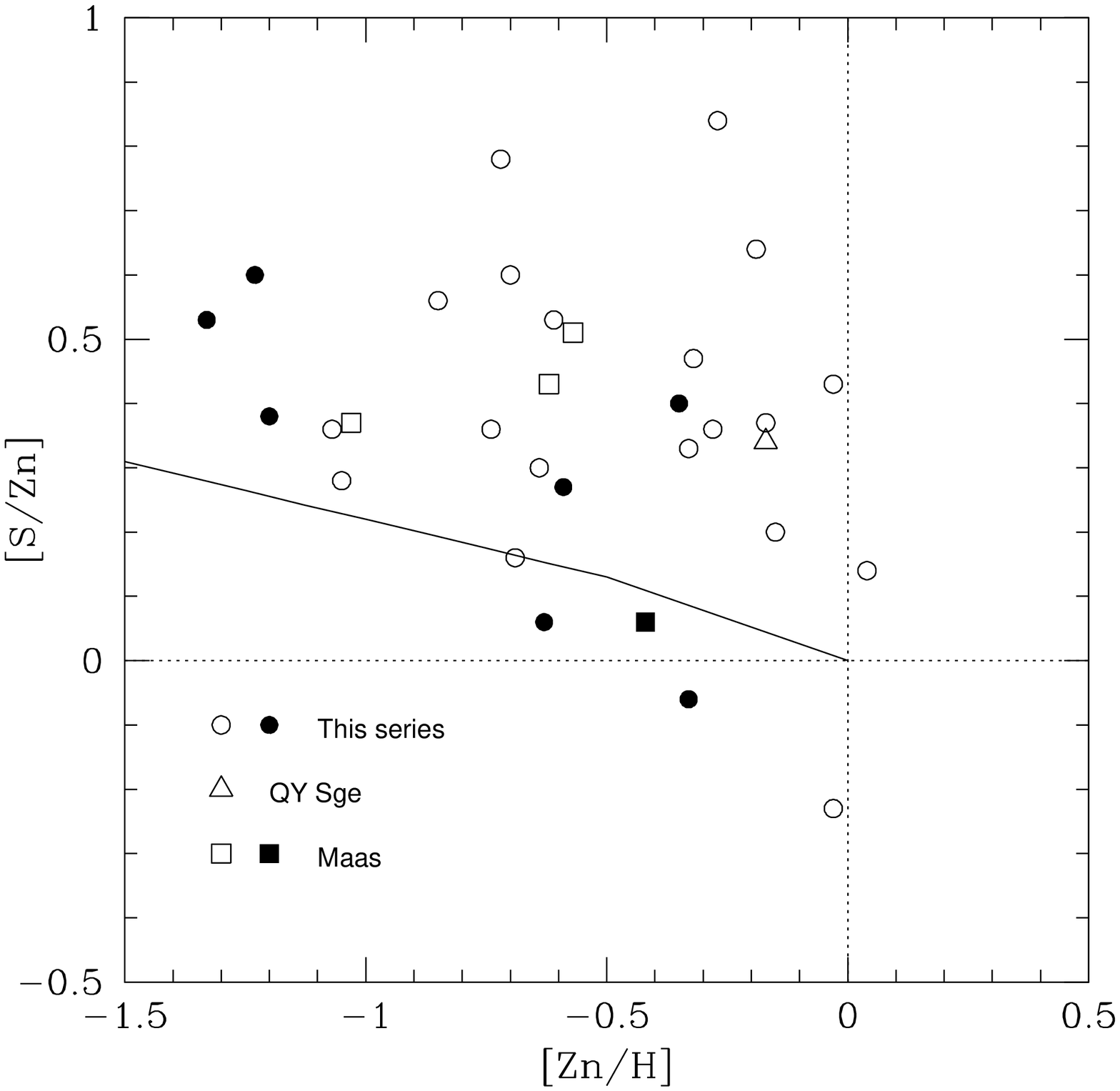}
\caption{Abundance ratio [S/Zn] versus abundance [Zn/H] for
RV Tauri and related variables.
 RV Tauri stars from our series of papers
are represented by circles. QY Sge (Rao et al. 2002) is shown by
the triangle. RV Tauri stars analysed by Maas et al. (2002) and
Maas (2003) are
shown by squares.
 Filled circles and squares denote high velocity stars. 
 The solid line shows the run of [S/Zn] with
[Zn/H] for unevolved stars (see text).  } 
\end{figure}
\clearpage

\begin{figure}
\plotone{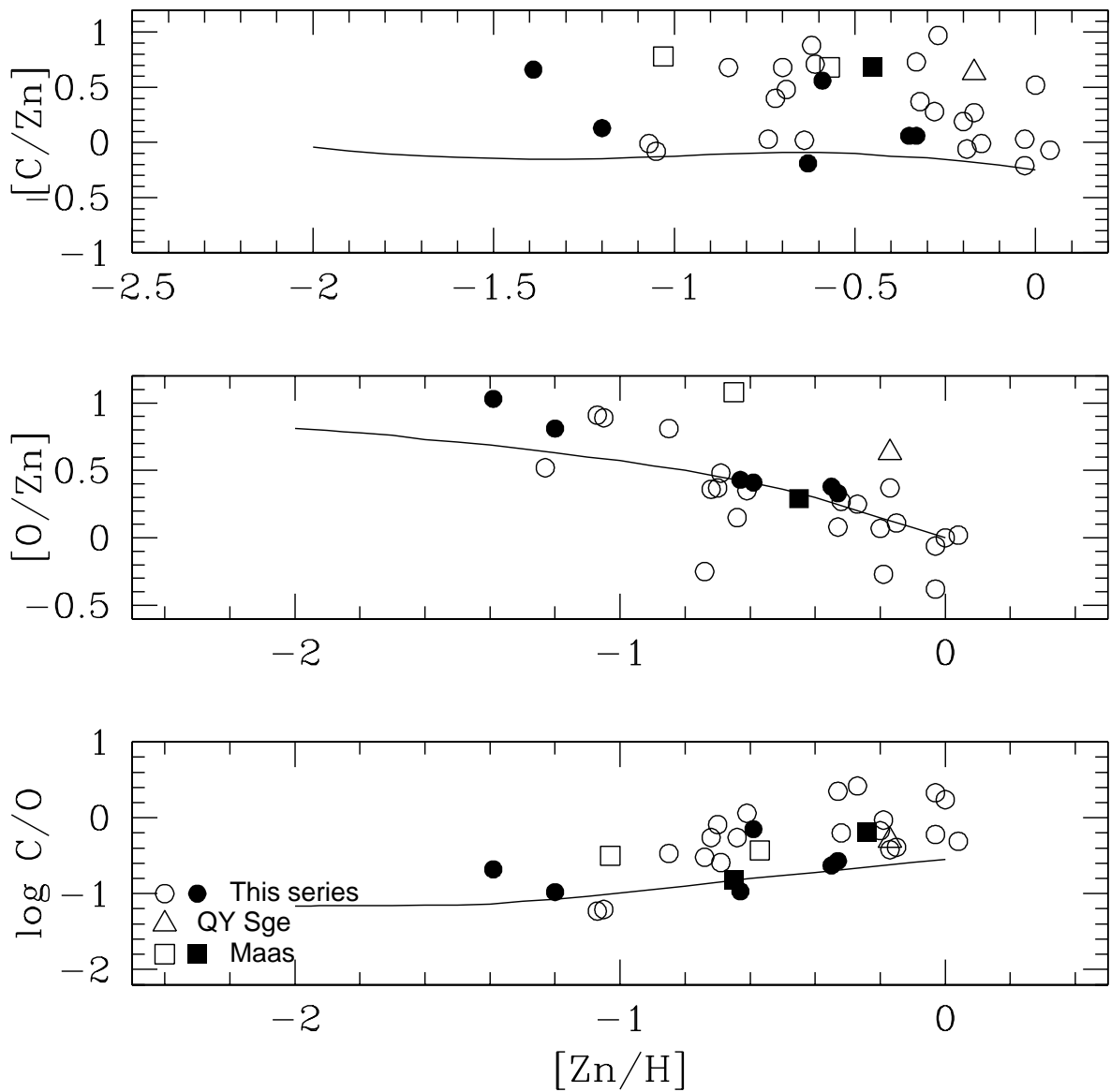}
\caption{The abundances [C/Zn] (top panel),
[O/Zn] (middle panel), and $\log$ C/O (bottom panel) as a function of the zinc
abundance [Zn/H] for RV Tauri and related variables.
 RV Tauri stars from our series of papers
are represented by circles. QY Sge (Rao et al. 2002) is shown by
the triangle. Stars analysed by Maas et al. (2002) and
Maas (2003) are
shown by squares.
 Filled circles and squares denote high velocity  stars.
The solid line shows the expected runs  with [Zn/H] for giants after the
first dredge-up
(see text).}
\end{figure}

\clearpage

\begin{figure}
\plotone{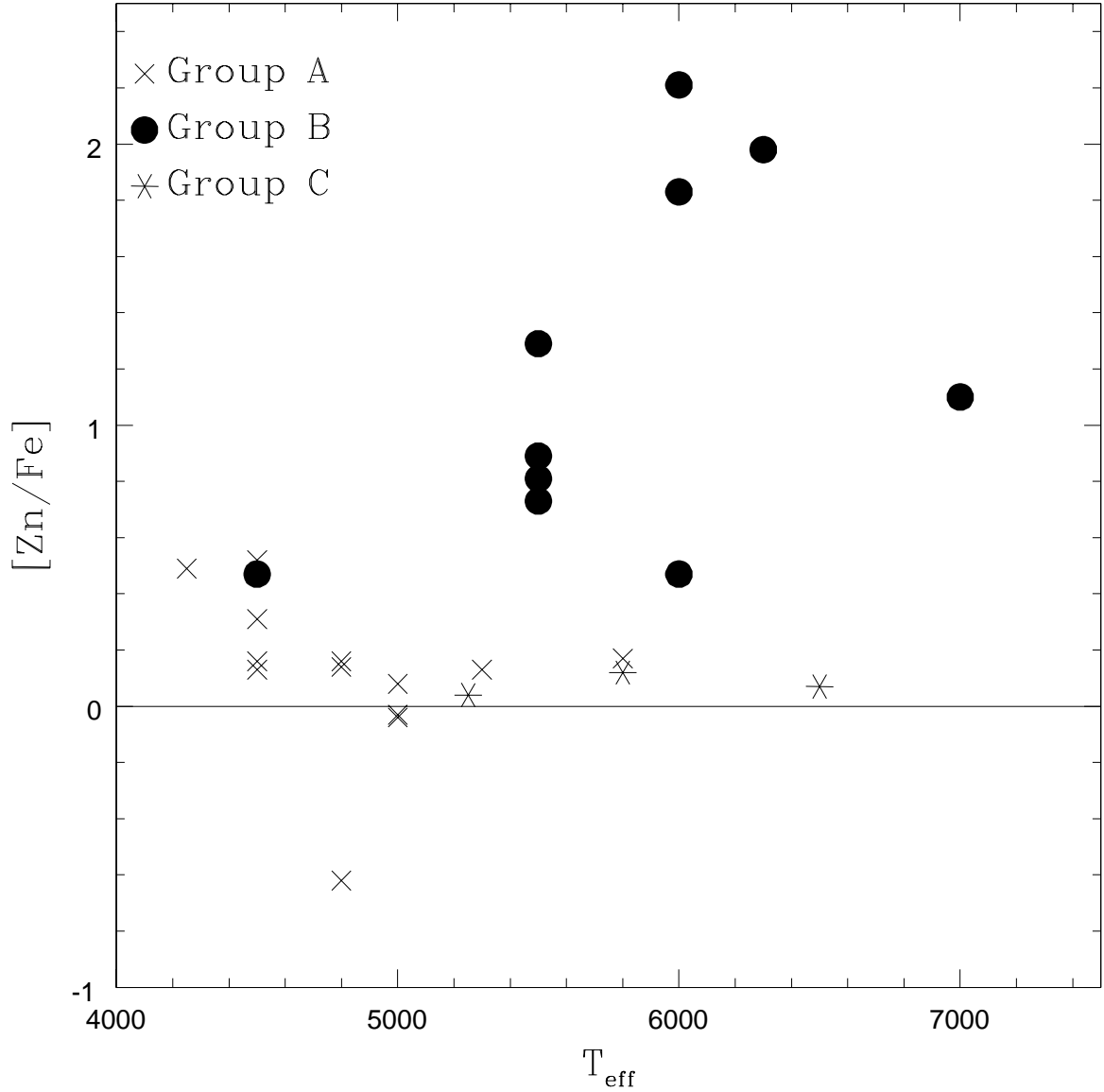}
\caption{ Abundance ratio [Zn/Fe] versus $T_{\rm eff}$ for RV Tauri stars.
Data points include the present work as well as from our earlier papers.
The ratio [Zn/Fe] increases towards hotter temperature indicating 
increasing severity of the dust-gas separation in hotter stars. 
to cooler ones. Symbols represent Preston spectroscopic groups A, B
and C. }
\end{figure}
\clearpage

\end{document}